\pdfoutput=1 

\documentclass[preprint,aps,prd,superscriptaddress,preprintnumbers,nofootinbib,longbibliography]{revtex4-1}

\usepackage[utf8]{inputenc}
\usepackage{graphicx}
\usepackage[T1]{fontenc}
\usepackage{amstext}
\usepackage{amsmath,amssymb,array}
\usepackage{babel}
\usepackage{color}
\usepackage{hyperref}
\usepackage{subcaption}
\usepackage{slashed}

\definecolor{smcpurple}{RGB}{90,70,120}

\begin{document}
	
	\title{Cosmic Evolution of the Standard Model Flavor Charges}
	
	\author{Chee Sheng Fong}
	\email{sheng.fong@ufabc.edu.br}
    \affiliation{Centro de Ciências Naturais e Humanas,
Universidade Federal do ABC, 09.210-170, Santo André, SP, Brazil}
	\author{Sergio Manuel Cubides Pérez}
	\email{sergio.manuel@ufabc.edu.br}
    \affiliation{Centro de Ciências Naturais e Humanas,
Universidade Federal do ABC, 09.210-170, Santo André, SP, Brazil}

\begin{abstract}

Since all the Standard Model (SM) parameters have been measured to precision at the percent level or smaller, one would aim to have a framework which describes \emph{baryogenesis} as \emph{precise} as possible, where the measured SM parameters are used as inputs. At high temperature before the electroweak symmetry breaking, the SM contains 15 approximate $U(1)$ symmetries with the associated flavor charges. To describe the evolutions of these flavor charges in a \emph{flavor-basis-covariant} manner such that physical observables are \emph{flavor-basis-independent}, a total of five density matrices in flavor spaces are required: three for quarks and two for leptons. Taking into account the quantum chromodynamics and electroweak sphaleron interactions together with all the Yukawa interactions, we obtain the complete flavor-covariant Boltzmann equations. 
By imposing chemical equilibrium in the lepton sector, we further derive a new \emph{effective quark-flavor-covariant} formalism which can be used when baryogenesis occurs through number asymmetry generation in the quark sector.
To verify the consistency of this complete formalism, we first apply it to leptogenesis scenarios, reproducing previous results which utilize the \emph{effective lepton-flavor-covariant} formalism up to small corrections due to the quark chemical equilibrium approximations used in the latter formalism.
Then, for the first time, we apply the complete formalism as well as the effective quark-flavor-covariant formalism to a cloistered baryogenesis scenario where number asymmetries are generated in the quark sector, showcasing the good agreement between the results from the two formalisms.
Finally, we release the first generic baryogenesis public code \texttt{BOLEH} (\texttt{BaryOn and Lepton charge Evolution in the Hot big bang}) where all the SM interactions are taken into account.

\end{abstract}

\maketitle
	
\newpage

\section{Introduction}

All the parameters of the Standard Model (SM) at some specific scales have now been measured to precision at percent level or better and have served as inputs for further experimental predictions. 
In cosmology, the SM will then serve as the benchmark particle physics model. Extrapolate back in time when the universe was smaller, denser and hotter, there were two main events that pointed to a universe in thermal equilibrium.
From the observation of cosmic microwave background (CMB) today, we know that the universe was in thermal equilibrium with temperature $T$ of eV scale during recombination when the CMB photons originated. 
From the measurements of light element abundances produced from the Big Bang Nucleosynthesis (BBN), we know that the universe should be radiation dominated with $T \sim$ MeV. It is plausible that the universe was in thermal equilibrium at higher temperature up to grand unified theory scale of $10^{16}$ GeV.\footnote{The current bound from nonobservation of inflationary gravitational wave by Planck and BICEP/Keck data~\cite{BICEP:2021xfz} implies an upper bound on the reheating temperature after inflation $T < 5.7 \times 10^{15}$ GeV from single field inflation.} 

Since the universe is observed to be \emph{matter-antimatter asymmetric}~\cite{Steigman:1976ev, Cohen:1997ac}, quantities of interest are the number density asymmetries of the SM fields i.e. the differences between number densities of SM particles and antiparticles. Before the electroweak (EW) symmetry breaking, the fields that can carry number density asymmetries are a scalar $SU(2)_L$ doublet Higgs $H$ together with
five types of fermionic fields that come in three flavors or families $(\alpha = 1,2,3)$: the $SU(2)_L$ doublets of quarks $Q_\alpha$ and leptons $\ell_\alpha$, the $SU(2)_L$ singlets of up-type quarks $U_\alpha$, down-type quarks $D_\alpha$, and charged leptons $E_\alpha$. To describe the evolutions of these number density asymmetries, one has to take into account all SM interactions that can change their numbers: QCD and EW sphaleron interactions~\cite{tHooft:1976rip,Moore:1997im} and all the SM Yukawa interactions. 
Due to conservation of hypercharge, the number density asymmetry in $H$ is not independent but can by expressed in terms of the number density asymmetries of the rest of the SM fermions, collectively known as the \emph{SM flavor charges}.

Since physical observables, like the total baryon and lepton charges, should be independent of flavor basis, one should work with flavor-covariant equations. 
In order to achieve this, the SM flavor charges $n_{\Delta \psi_\alpha}$ of five types of SM fermion fields $\psi_\alpha$ should be described by matrices $(n_{\Delta\psi})_{\alpha\beta}$ in their respective flavor spaces~\cite{Sigl:1992fn}. Under a change of flavor basis $\psi \to V \psi$ where $V$ is a unitary matrix, we have $n_{\Delta \psi} \to V n_{\Delta\psi} V^\dagger$ while the total flavor charge is given by ${\rm Tr}(n_{\Delta\psi})$ which is independent of flavor basis.
Ref.~\cite{Fong:2021xmi} shows that in order to describe evolution of lepton flavor charges in a flavor-basis-independent manner, we need to follow the evolutions of both $n_{\Delta \ell}$ and $n_{\Delta E}$ while the effects of quark Yukawa interactions can be captured effectively through temperature-dependent coefficients, obtained by assuming quark chemical equilibrium at a given temperature. In this work, we first generalize the study to include also the evolutions of three quark flavor charges $n_{\Delta Q}$, $n_{\Delta U}$ and $n_{\Delta D}$. One will then be able to describe the evolutions of all the 15 SM flavor charges~\cite{Fong:2020fwk} taking into account their flavor correlations, using only the measured SM parameters as inputs. 
Then, we further derive a new \emph{effective quark-flavor-covariant} formalism which is applicable to baryogenesis mechanism that generates number asymmetries in the quark sector, while the lepton sector is assumed to be in chemical equilibrium at a given temperature. 

To verify the complete formalism, we first apply it to the well-established leptogenesis scenarios~\cite{Fukugita:1986hr,Fong:2012buy} and compare the results against those obtained from the effective lepton-flavor-covariant formalism of ref.~\cite{Fong:2020fwk}. 
As a demonstration of the utility of the complete formalism as well as the effective quark-flavor-covariant formalism, we apply them to a cloistered baryogenesis scenario in which asymmetry generation occurs in the quark charges~\cite{AristizabalSierra:2013lyx}.
Together with this work, we release the first generic baryogenesis public code \texttt{BOLEH} (\texttt{BaryOn and Lepton charge Evolution in the Hot big bang}) implementing the three formalisms: (i) the complete formalism including all the SM interactions, (ii) the effective quark-flavor-covariant formalism and (iii) the effective lepton-flavor-covariant formalism.

This article is organized as follows. In Section~\ref{sec:SM_BEs}, we first introduce the complete SM flavor-covariant Boltzmann equations and then in Section~\ref{sec:effQ}, we derive the effective quark-flavor-covariant formalism.  
In Section~\ref{sec:leptogenesis}, we apply the complete formalism as well as the effective lepton-flavor-covariant formalism to the type-I and type-II leptogenesis scenarios.
For completeness, the relevant equations of the effective lepton-flavor-covariant formalism are collected in Appendix~\ref{app:lepton_flavor_charges}. In Section~\ref{sec:cloistered}, we apply the complete and the effective quark-flavor-covariant formalisms to a cloistered baryogenesis model where quarks do not achieve chemical equilibrium during baryogenesis. Finally, we conclude in Section~\ref{sec:conclusions}. In Appendix~\ref{app:code}, we describe the generic baryogenesis code \texttt{BOLEH}.

\section{Standard Model Flavor-covariant Boltzmann Equations}\label{sec:SM_BEs}

\subsection{Standard Model Flavor Charges}

Assuming a radiation-dominated universe, all the SM fields will be in thermal equilibrium due to fast gauge interactions, and a particle $\psi$ and its antiparticle $\bar \psi$ can be described by generalized equilibrium phase space distributions at temperature $T$ as
\begin{eqnarray}
	\left(f_{\psi,\mathbf{p}}\right)_{\alpha\beta} & = & \left[e^{\frac{{\cal E}_{\psi}-(\mu_{\psi})_{\alpha\beta}}{T}}+\xi_{\psi}\right]^{-1},
	\;\;\;\;\;
	\left(f_{\bar{\psi},\mathbf{p}}\right)_{\alpha\beta} = \left[e^{\frac{{\cal E}_{\psi}+(\mu_{\psi})_{\alpha\beta}}{T}}+\xi_{\psi}\right]^{-1},
	\label{eq:f_i_ibar}
\end{eqnarray}
where $\xi_{\psi} = 1(-1)$ for fermion (boson), ${\cal E}_{\psi}= \sqrt{|\mathbf{p}|^2 + m_\psi^2}$ is the energy of $\psi$ with mass $m_\psi$, and $\mu_{\psi}$ is the generalized chemical potential where $\alpha,\beta$ denote possible flavor indices. In the SM, the five types of fermion fields $\psi = \{Q,U,D,\ell,E\}$ are described by five $3\times 3$ \emph{Hermitian} matrices of $(\mu_\psi)_{\alpha\beta}$ while the Higgs is described by a $\mu_H$. Expanding the phase space distributions to linear order in $\mu/T$ and integrating over 3-momentum including the additional gauge degrees of freedom $g_\psi$, we can define the number density asymmetry as
\begin{eqnarray}
	\left(n_{\Delta \psi}\right)_{\alpha\beta} & \equiv & 
	g_\psi \int \frac{d^3 p}{(2\pi)^3} 
	\left[\left(f_{\psi,\mathbf{p}}\right)_{\alpha\beta} 
	- \left(f_{\bar \psi,\mathbf{p}}\right)_{\alpha\beta}\right]
	=\frac{T^{2}}{6}g_{\psi}\zeta_{\psi}\left(\mu_{\psi}\right)_{\alpha\beta},
	\label{eq:density_asymmetry_chempot}
\end{eqnarray}
where
\begin{eqnarray}
	\zeta_{\psi} & \equiv & \frac{6}{\pi^{2}}\int_{m_{\psi}/T}^{\infty}dxx\sqrt{x^{2}-m_{\psi}^{2}/T^{2}}\frac{e^{x}}{\left(e^{x} + \xi_{\psi}\right)^{2}}. \label{eq:zeta}
\end{eqnarray}
In the SM, we will take $m_{\psi}=0$ such that $\zeta_{\psi}=1(2)$ for fermion (boson). 

Under a unitary transformation in the flavor space of a type of the SM fermion $\psi \to V_\psi \psi$, we can define such that eq.~\eqref{eq:density_asymmetry_chempot} transforms as~\cite{Sigl:1992fn,Fong:2021xmi}
\begin{equation}
	\left(n_{\Delta \psi}\right)_{\alpha\beta} \to V_\psi \left(n_{\Delta \psi}\right)_{\alpha\beta} V_\psi^\dagger.
	\label{eq:flavor_transformation}
\end{equation}
The total number density asymmetry of a type of the SM fermion is given by ${\rm Tr}(n_{\Delta \psi})$ which is \emph{independent} of flavor basis.

Since the comoving entropy is conserved in radiation-dominated universe, it is convenient to normalize eq.~\eqref{eq:density_asymmetry_chempot} by the cosmic entropy density $s=\frac{2\pi^{2}}{45}g_{\star}T{{}^3}$ where $g_{\star}$ denotes the effective relativistic degrees of freedom to obtain
\begin{eqnarray}
	\left(Y_{\Delta \psi}\right)_{\alpha\beta} & \equiv & \frac{\left(n_{\Delta \psi}\right)_{\alpha\beta}}{s}\equiv Y^{{\rm nor}}g_{\psi}\zeta_{\psi}\frac{2(\mu_{\psi})_{\alpha\beta}}{T},
	\label{eq:Y_mu}
\end{eqnarray}
with $Y^{\rm nor} \equiv 15/(8\pi^2 g_\star)$. 
For instance, for the up-type quark, we have $g_U = 3$ and $\xi_U = 1$ giving a $3\times 3$ up-type quark density matrix $Y_{\Delta U} = 6 Y^{\rm nor} \mu_U/T$ while for the Higgs, we have $g_H = 2$ and $\xi_H = 2$ giving a Higgs charge $Y_{\Delta H} = 8 Y^{\rm nor} \mu_H/T$.
Due to hypercharge conservation, $Y_{\Delta H}$ is not an independent charge but is related to the other SM flavor charges as follows 
\begin{eqnarray}
	Y_{\Delta H} & = & 
	\frac{1}{3}{\rm Tr}\left[-Y_{\Delta Q}-4Y_{\Delta U}+2Y_{\Delta D}\right]
	+{\rm Tr}\left[Y_{\Delta\ell}+2Y_{\Delta E}\right]
	- 2\sum_i q^Y_{\phi_i} Y_{\Delta \phi_i},
	\label{eq:YH}
\end{eqnarray}
where we include the last term to take into account possible beyond the SM scalar fields $\phi_i$ which are in kinetic equilibrium and carry nonzero hypercharge $q^Y_\phi$. The additional gauge multiplicity $g_{\phi_i}$ and the mass of $\phi_i$ in $\zeta_{\phi_i}$ have been included in the definition of $Y_{\Delta\phi_i}$. 
The expression \eqref{eq:YH} holds also if $\phi_i$ are fermions but to consistently introduce new hypercharge-carrying fermions, one will have to take into account gauge anomaly cancellations and therefore, we will focus only on the case of scalars.

Finally, for a radiation-dominated, homogeneous and isotropic universe, the Hubble expansion rate is given by the Friedmann equation ${\cal H}=1.66\sqrt{g_{\star}}T^{2}/M_{{\rm Pl}}$ with $M_{\rm Pl} = 1.22\times 10^{19}$ GeV and we can trade time $t$ for monotonously decreasing temperature $T$ to write the evolution of a flavor charge in time as
\begin{equation}
	\frac{dn_{\Delta \psi}}{dt} + 3 {\cal H}n_{\Delta \psi} = s\frac{dY_{\Delta\psi}}{dt} = s{\cal H}z\frac{dY_{\Delta\psi}}{dz},
\end{equation}
where $z \equiv M_{\rm ref}/T$ with $M_{\rm ref}$ an arbitrary reference mass scale.

\subsection{Standard Model Interactions}

In the evolution of the SM flavor charges, the SM Yukawa interactions play a crucial role since they determine the effective conserved charges in the early Universe~\cite{Fong:2015vna,Fong:2020fwk}.
They are described by
\begin{eqnarray}
	-{\cal L} \supset (y_U)_{\alpha \beta} \overline {U_\alpha} Q_\beta \epsilon H + (y_D)_{\alpha \beta} \overline {D_\alpha} Q_\beta H^* + (y_E)_{\alpha \beta} \overline {E_\alpha} \ell_\beta H^* + {\rm H.c.},\label{eq:Yukawa_terms}
\end{eqnarray}
where $y_U$, $y_D$ and $y_E$ are Yukawa couplings with flavor indices $\alpha,\beta = 1,2,3$ and $\epsilon$ is the total antisymmetric tensor in the $SU(2)_L$ space.\footnote{We have suppressed the left- and right-handed projectors: under Lorentz group, $Q,\ell$ transform as left-handed fields while $U,D,E$ transform as right-handed fields.} In the limit where the Yukawa couplings go to zero, there are 16 conserved charges, each associated to the $U(1)_\psi$ transformation of the corresponding SM field: 1 Higgs and 15 fermions. Turning on the Yukawa interactions, the remaining symmetries are $U(1)_Y$ hypercharge gauge symmetry and 4 global $U(1)$'s: the total baryon number $B$ and three lepton flavor numbers $L_\alpha$ defined in the basis where $y_E$ is diagonal. Using $q^X_\psi$ to denote the charge of $\psi$ under a $U(1)_X$, we will adopt the convention $6q^Y_Q = 3q^Y_U/2 = -3q^Y_D = -2q^Y_\ell = -q^Y_E = 2q^Y_H = 1$, $q^B_Q = q^B_U = q^B_L =1/3$, $q^{L_\alpha}_{\ell_\alpha} = q^{L_\alpha}_{E_\alpha} = 1$ and zero otherwise. Due to the observed structures of $y_U$, $y_D$ and $y_E$, effective $U(1)$ symmetries arise at high temperature $T$ when interactions in some directions in flavor space go out-of-equilibrium and the resulting effective charges can play interesting roles in baryogenesis~\cite{Fong:2015vna,Domcke:2020quw,Fong:2020fwk}.

In the SM, due to the Adler-Bell-Jackiw anomaly, there are nonperturbative interactions that break the effective symmetries.
Quark chiral symmetries are anomalous under $SU(3)_c$ and are explicitly broken by instanton-induced effective operator~\cite{Moore:1997im}
\begin{eqnarray}
	{\cal O}_{SU(3)} = \prod_\alpha Q_\alpha Q_\alpha U^c_\alpha D^c_\alpha.
\end{eqnarray}
Analogously, baryon $B$ and lepton $L$ are anomalous under $SU(2)_L$ and are explicitly broken by~\cite{tHooft:1976rip}
\begin{eqnarray}
	{\cal O}_{SU(2)} = \prod_\alpha Q_\alpha Q_\alpha Q_\alpha \ell_\alpha,
\end{eqnarray}
while the linear combinations $B_\alpha-L_\alpha$ (one for each flavor) are exactly conserved.
At high temperature, these operators are unsuppressed and the $SU(N)$ sphaleron rate per volume are determined by refs.~\cite{Moore:1997im,Moore:2000mx,Moore:2000ara,Garbrecht:2014kda} to be
\begin{eqnarray}
	\gamma_{N} & = & \left(8.24\pm0.10\right)\left(\frac{N}{2}\right)^{2}\frac{g_{N}^{2}T^{2}}{m_{N}^{2}}\left(\ln\frac{m_{N}}{g_{N}^{2}T}+3.041\right)\alpha_{N}^{5}T^{4},
\end{eqnarray}
where $g_{N}$ is the $SU(N)$ gauge coupling with $\alpha_{N}\equiv g_{N}^{2}/\left(4\pi\right)$
and $m_{N}^{2}$ is the Debye mass squared given by $m_{2}^{2}=11g_{2}^{2}T^{2}/6$
and $m_{3}^{2}=2g_{3}^{2}T^{2}$. This gives, for the SM $SU(2)_L$ and $SU(3)_c$, respectively
\begin{eqnarray}
	\gamma_{\rm EW} & \equiv & \gamma_2 = \left(13.7+4.49\ln\frac{1.35}{g_{2}}\right)\alpha_{2}^{5}T^{4},\\
	\gamma_{\rm QCD} & \equiv & \gamma_3 = \left(95.1+31.3\ln\frac{1.41}{g_{3}}\right)\alpha_{3}^{5}T^{4},
\end{eqnarray}
where we have used the one-loop renormalization group equation (RGE) running of the SM gauge couplings at energy scale $\mu=2\pi T$ with $T = 10^{12}$ GeV for EW and $T = 10^{13}$ GeV for QCD.

Considering the Yukawa interactions in eq.~\eqref{eq:Yukawa_terms} up to the leading order as well as the sphaleron interactions, we have \cite{Garbrecht:2014kda,Fong:2021xmi}
\begin{eqnarray}
	s{\cal H}z\frac{dY_{\Delta Q}}{dz} & = & -3{\cal C}_{{\rm EW}}-2{\cal C}_{{\rm QCD}} \nonumber \\
	& & -\frac{\gamma_{U}}{2Y^{{\rm nor}}}\left\{ y_{U}^{\dagger}y_{U},\frac{Y_{\Delta Q}}{g_{Q}\zeta_{Q}}\right\} -\frac{\gamma_{U}}{Y^{{\rm nor}}}y_{U}^{\dagger}y_{U}\frac{Y_{\Delta H}}{g_{H}\zeta_{H}}+\frac{\gamma_{U}}{Y^{{\rm nor}}}y_{U}^{\dagger}\frac{Y_{\Delta U}}{g_{U}\zeta_{U}}y_{U}\nonumber \\
	&  & -\frac{\gamma_{D}}{2Y^{{\rm nor}}}\left\{ y_{D}^{\dagger}y_{D},\frac{Y_{\Delta Q}}{g_{Q}\zeta_{Q}}\right\} +\frac{\gamma_{D}}{Y^{{\rm nor}}}y_{D}^{\dagger}y_{D}\frac{Y_{\Delta H}}{g_{H}\zeta_{H}}+\frac{\gamma_{D}}{Y^{{\rm nor}}}y_{D}^{\dagger}\frac{Y_{\Delta D}}{g_{D}\zeta_{D}}y_{D},
	\label{eq:BEQ}\\
	s{\cal H}z\frac{dY_{\Delta U}}{dz} & = & {\cal C}_{{\rm QCD}} -\frac{\gamma_{U}}{2Y^{{\rm nor}}}\left\{ y_{U}y_{U}^{\dagger},\frac{Y_{\Delta U}}{g_{U}\zeta_{U}}\right\} +\frac{\gamma_{U}}{Y^{{\rm nor}}}y_{U}y_{U}^{\dagger}\frac{Y_{\Delta H}}{g_{H}\zeta_{H}}+\frac{\gamma_{U}}{Y^{{\rm nor}}}y_{U}\frac{Y_{\Delta Q}}{g_{Q}\zeta_{Q}}y_{U}^{\dagger},
	\label{eq:BEU}\\
	s{\cal H}z\frac{dY_{\Delta D}}{dz} & = & {\cal C}_{{\rm QCD}} -\frac{\gamma_{D}}{2Y^{{\rm nor}}}\left\{ y_{D}y_{D}^{\dagger},\frac{Y_{\Delta D}}{g_{D}\zeta_{D}}\right\} -\frac{\gamma_{D}}{Y^{{\rm nor}}}y_{D}y_{D}^{\dagger}\frac{Y_{\Delta H}}{g_{H}\zeta_{H}}+\frac{\gamma_{D}}{Y^{{\rm nor}}}y_{D}\frac{Y_{\Delta Q}}{g_{Q}\zeta_{Q}}y_{D}^{\dagger},
	\label{eq:BED}\\
	s{\cal H}z\frac{dY_{\Delta\ell}}{dz} & = & -{\cal C}_{{\rm EW}} -\frac{\gamma_{E}}{2Y^{{\rm nor}}}\left\{ y_{E}^{\dagger}y_{E},\frac{Y_{\Delta\ell}}{g_{\ell}\zeta_{\ell}}\right\} +\frac{\gamma_{E}}{Y^{{\rm nor}}}y_{E}^{\dagger}y_{E}\frac{Y_{\Delta H}}{g_{H}\zeta_{H}}+\frac{\gamma_{E}}{Y^{{\rm nor}}}y_{E}^{\dagger}\frac{Y_{\Delta E}}{g_{E}\zeta_{E}}y_{E},
	\label{eq:BEell}\\
	s{\cal H}z\frac{dY_{\Delta E}}{dz} & = & -\frac{\gamma_{E}}{2Y^{{\rm nor}}}\left\{ y_{E}y_{E}^{\dagger},\frac{Y_{\Delta E}}{g_{E}\zeta_{E}}\right\} -\frac{\gamma_{E}}{Y^{{\rm nor}}}y_{E}y_{E}^{\dagger}\frac{Y_{\Delta H}}{g_{H}\zeta_{H}}+\frac{\gamma_{E}}{Y^{{\rm nor}}}y_{E}\frac{Y_{\Delta\ell}}{g_{\ell}\zeta_{\ell}}y_{E}^{\dagger},\label{eq:BEE}
\end{eqnarray}
where $g_{Q}=2g_{U}=2g_{D}=6$, $g_{\ell}=g_{H}=2g_{E}=2$, $\zeta_Q = \zeta_U = \zeta_D = \zeta_{\ell}=\zeta_{E}=1=\zeta_{H}/2$ and
\begin{eqnarray}
	{\cal C}_{{\rm EW}} & \equiv & \frac{\gamma_{{\rm EW}}}{4Y^{{\rm nor}}}\left(\frac{{\rm Tr}Y_{\Delta\ell}}{g_{\ell}\zeta_{\ell}}+3\frac{{\rm Tr}Y_{\Delta Q}}{g_{Q}\zeta_{Q}}\right)I_{3\times3}, \\
	{\cal C}_{{\rm QCD}} & \equiv & \frac{\gamma_{{\rm QCD}}}{6Y^{{\rm nor}}}\left(2\frac{{\rm Tr}Y_{\Delta Q}}{g_{Q}\zeta_{Q}}-\frac{{\rm Tr}Y_{\Delta U}}{g_{U}\zeta_{U}}-\frac{{\rm Tr}Y_{\Delta D}}{g_{D}\zeta_{D}}\right)I_{3\times3}.
\end{eqnarray}
The rates of Yukawa interactions are captured by
$\gamma_U$, $\gamma_D$ and $\gamma_E$ (with Yukawa couplings factored out) which depend on the gauge couplings and mainly the top quark Yukawa coupling. Using the results of refs.~\cite{Garbrecht:2013bia,Garbrecht:2014kda} and considering one-loop RGE running of the SM couplings, we obtain the following fitting functions
\begin{eqnarray}
	\frac{\gamma_{U,D}}{T^4} &=& -8.2\times10^{-6} L_T^3 + 3.0\times10^{-4} L_T^2 - 4.0\times10^{-3} L_T + 3.0\times10^{-2}, \\
	\frac{\gamma_{E}}{T^4} &=& -5.0\times10^{-7} L_T^3 + 2.1\times10^{-5} L_T^2 - 3.2\times10^{-4} L_T + 6.6\times10^{-3},
\end{eqnarray}
where $L_T \equiv \log_{10} (T/1\, {\rm GeV})$. They are accurate to subpercent level for temperature in the range $T = (100,10^{15})$ GeV.

The set of Boltzmann equation \eqref{eq:BEQ}--
\eqref{eq:BEE} are \emph{flavor-covariant}: 
under a change of basis $\psi \to V_\psi \psi$, the terms on the right transform covariantly consistent with eq.~\eqref{eq:flavor_transformation}. 
The basis-independent total baryon $Y_B$ and lepton $Y_L$ charges follow
\begin{eqnarray}
	s {\cal H}z \frac{dY_B}{dz} & = & \frac{1}{3}s {\cal H}z \frac{d}{dz} {\rm Tr}(Y_{\Delta Q} + Y_{\Delta U} + Y_{\Delta D}) = - 3 {\cal C}_{{\rm EW}}, \\
	s {\cal H}z \frac{dY_L}{dz} &=& s {\cal H}z \frac{d}{dz} {\rm Tr}(Y_{\Delta \ell} + Y_{\Delta E}) = - 3 {\cal C}_{{\rm EW}}\label{LeptonAsym},
\end{eqnarray}
where all the Yukawa interactions drop out as they do not violate $B$ and $L$. We can further verify that $Y_{B-L} \equiv Y_B - Y_L$ is conserved since $B-L$ is preserved by the EW sphaleron interaction.

Nonzero $Y_{B-L}$ can be generated by introducing \emph{new physics} processes which violate $B-L$, as source terms in any of the Boltzmann equations \eqref{eq:BEQ}--\eqref{eq:BEE}.
When the EW sphaleron interaction freezes out at $T = 132$ GeV~\cite{DOnofrio:2014rug}, the final $B$ asymmetry is given by~\cite{Fong:2015vna,Babu:2024ahk}
\begin{eqnarray}
	Y_{B}^{\rm final} & = & 0.315Y_{B-L}\Big|_{T = 132\,{\rm GeV}}.\label{eq:YB_YB-L}
\end{eqnarray}
The value obtained above from a barygoenesis mechanism should be compared against the observed value: 
$Y_B^{\rm obs} = (8.70 \pm 0.06) \times 10^{-11}$~\cite{Planck:2018vyg}.

\section{Effective evolution of baryon charges}
\label{sec:effQ}
New physics that generates asymmetries only in the quark sector ($Y_{\Delta Q}, Y_{\Delta U}, Y_{\Delta D}$) will result in deviation from chemical equilibrium in the quark sector. In this case, it is a good approximation to impose chemical equilibrium in the lepton sector when the lepton Yukawa interactions get into equilibrium at the corresponding temperatures~\cite{Fong:2020fwk}
\begin{equation}
    T_\tau = 4 \times 10^{11}\,{\rm GeV},\quad
    T_\mu = 10^{9}\,{\rm GeV},\quad
    T_e = 3 \times 10^{4}\,{\rm GeV},
\end{equation}
when the number of right-handed $\tau = E_3$, $\mu = E_2$ and $e = E_1$ become conserved, respectively. 
While these transitions are captured exactly by the complete formalism introduced in the previous section, here, their effects will be captured by some temperature-dependent coefficients that we will derive next.

First of all, we define the charge matrix 
\begin{equation}
Y_{\tilde \Delta QL} \equiv \frac{1}{3}\left(Y_{\Delta Q}-Y_{L} I_{3\times3}\right),
\end{equation}
which is conserved by the EW sphaleron interactions.
The normalization factor $1/3$ is chosen such that the trace of the effective charge matrix, together with the singlet quark asymmetries, directly reproduces $Y_{B-L}$,
\begin{equation}
        Y_{B-L}={\rm Tr}Y_{\tilde\Delta QL}+\frac{1}{3}{\rm Tr}Y_{\Delta U}+\frac{1}{3}{\rm Tr}Y_{\Delta D}.
\end{equation}
Notice that we only have to follow the evolution of $Y_{\tilde \Delta QL}$ in place of the evolutions of $Y_{\Delta \ell}$ and $Y_{\Delta E}$, thereby reducing the variables of total system by $2 \times 9 = 18$, corresponding to that of the two Hermitian matrices $Y_{\Delta \ell}$ and $Y_{\Delta E}$.

From eqs.~\eqref{eq:BEQ} and \eqref{eq:BEell}, we obtain the Boltzmann equation for $Y_{\tilde \Delta QL}$
\begin{eqnarray}
s{\cal H}z\frac{dY_{\tilde\Delta QL}}{dz} &=& 
  -\frac{2}{3}{\cal C}_{{\rm QCD}} 
	 -\frac{\gamma_{U}}{6Y^{{\rm nor}}}\left\{ y_{U}^{\dagger}y_{U},\frac{Y_{\Delta Q}}{g_{Q}\zeta_{Q}}\right\} -\frac{\gamma_{U}}{3Y^{{\rm nor}}}y_{U}^{\dagger}y_{U}\frac{Y_{\Delta H}}{g_{H}\zeta_{H}}+\frac{\gamma_{U}}{3Y^{{\rm nor}}}y_{U}^{\dagger}\frac{Y_{\Delta U}}{g_{U}\zeta_{U}}y_{U} \nonumber \\
	&  & -\frac{\gamma_{D}}{6Y^{{\rm nor}}}\left\{ y_{D}^{\dagger}y_{D},\frac{Y_{\Delta Q}}{g_{Q}\zeta_{Q}}\right\} +\frac{\gamma_{D}}{3Y^{{\rm nor}}}y_{D}^{\dagger}y_{D}\frac{Y_{\Delta H}}{g_{H}\zeta_{H}}+\frac{\gamma_{D}}{3Y^{{\rm nor}}}y_{D}^{\dagger}\frac{Y_{\Delta D}}{g_{D}\zeta_{D}}y_{D},
\end{eqnarray}
where the EW sphaleron term drops out.

Together with the Boltzmann equations for quark singlets \eqref{eq:BEU} and \eqref{eq:BED}, we will need to rewrite $Y_{\Delta Q}$ and $Y_{\Delta H}$ in terms of $Y_{\tilde \Delta QL}$, $Y_{\Delta U}$ and $Y_{\Delta D}$ to form a closed set of Boltzmann equations.
This can be achieved by relating the asymmetry of a particle species $Y_{\Delta i}$ to charges $Y_{\Delta x}$ through~\cite{Fong:2015vna}
\begin{equation}
    Y_{\Delta i}=\sum_xg_i\zeta_i\sum_y q^y_i(J^{-1})_{yx}Y_{\Delta x},\label{eq:cahrgerel}
\end{equation}
where $x, y$ run through all effective charges of the system at a particular temperature regime and the $J$ matrix is defined by
\begin{equation}
    J_{xy}=\sum_ig_i\zeta
    _iq^x_{i}q^y_i,
\end{equation}
with $i$ sums over all particle species.

Including additional scalar fields $\phi_i$ with hypercharge $q_{\phi_i}^Y$, we obtain the following relations through eq.~\eqref{eq:cahrgerel}
\begin{eqnarray}
Y_{\Delta Q} &=& 3Y_{\tilde \Delta QL}+c_{Q_1}(T){\rm Tr}Y_{\tilde\Delta}I_{3\times3}+c_{Q_2}(T)Y_R I_{3\times3},\\
Y_{\Delta H} &=& c_{H_1}(T){\rm Tr}Y_{\tilde\Delta QL}+c_{H_2}(T) Y_R,
\end{eqnarray}
where $c_{Q_1}(T)$, $c_{Q_2}(T)$, $c_{H_1}(T)$ and $c_{H_2}(T)$ are some temperature-dependent coefficients and
and we have defined
\begin{equation}
    Y_R \equiv 2{\rm Tr}Y_{\Delta U}-{\rm Tr}Y_{\Delta D}+3\sum_iq_{\phi_i}^YY_{\Delta\phi_i}.
\end{equation}
Note that $Y_{\Delta\phi_i}$ takes into account the additional gauge multiplicity $g_{\phi_i}$ and the mass of $\phi_i$ in $\zeta_{\phi_i}$ through eq.~\eqref{eq:Y_mu}.

At temperature $T > T_B \equiv 2.3 \times 10^{12}\,{\rm GeV}$ when the EW sphaleron interactions are out of equilibrium, the total baryon and lepton numbers are conserved\footnote{Since $B-L$ is conserved in the SM, they are not independent conditions and one only has to impose either baryon or lepton number conservation.}. 
In this temperature regime, there are 14 effective charges: besides the 9 quark charges $(Y_{\tilde\Delta QL})_{\alpha\alpha}$, $(Y_{\Delta U})_{\alpha\alpha}$, $(Y_{\Delta D})_{\alpha\alpha}$ with $\alpha = 1,2,3$, we also have $U(1)_{E_3}$, $U(1)_{E_2}$, $U(1)_{E_1}$, hypercharge $U(1)_Y$ and the total baryon $U(1)_B$ (or lepton $U(1)_L$) number, in which the last five charges are set to zero, in the absence of preexisting asymmetries.
As $T$ drops below $T_B$, $T_\tau$, $T_\mu$, and $T_e$, respectively, the corresponding charges $U(1)_B$, $U(1)_{E_3}$, $U(1)_{E_2}$, and $U(1)_{E_1}$ cease to be conserved. In each case, we solve eq.~\eqref{eq:cahrgerel} for $c_{Q_1}$, $c_{Q_2}$, $c_{H_1}$ and $c_{H_2}$, and the results are summarized in Table~\ref{tab:effective_BFC}.

The transitions between the temperature regime of the coefficients are well described by the following exponential parametrization
\begin{eqnarray}
    c_{Q_1}(T) &=& -\frac{3}{4}\left(1-e^{-T_B/T}\right) -\left(-\frac{3}{4}+\frac{8}{11}\right)\left(1-e^{-T_\tau/T}\right)-\left(-\frac{8}{11}+\frac{5}{7}\right)\left(1-e^{-T_\mu/T}\right)\nonumber\\
    & & -\left(-\frac{5}{7}+\frac{12}{17}\right)\left(1-e^{-T_e/T}\right),\\
    c_{Q_2}(T) &=& \frac{1}{33}\left(1-e^{-T_\tau/T}\right)-\left(\frac{1}{33}-\frac{1}{21}\right)\left(1-e^{-T_\mu/T}\right)-\left(\frac{1}{21}-\frac{1}{17}\right)\left(1-e^{-T_e/T}\right),\\
    c_{H_1}(T) &=& -1-\left(-1+\frac{10}{11}\right)\left(1-e^{-T_\tau/T}\right)-\left(-\frac{10}{11}+\frac{6}{7}\right)\left(1-e^{-T_\mu/T}\right)\nonumber\\
    & &-\left(-\frac{6}{7}+\frac{14}{17} \right)\left(1-e^{-T_e/T}\right),\\
    c_{H_2}(T) &=& -\frac{2}{3}-\left(-\frac{2}{3}+\frac{6}{11}\right)\left(1-e^{-T_\tau/T}\right)-\left(-\frac{6}{11}+\frac{10}{21}\right)\left(1-e^{-T_\mu/T}\right)\nonumber\\
    & & -\left(-\frac{10}{21}+\frac{22}{51}\right)\left(1-e^{-T_e/T}\right).
\end{eqnarray}
We will denote this formalism as effective quark-flavor-covariant (effQ) formalism.

\begin{table}
\begin{tabular}{|c|c|c|c|c|c|}
\hline 
Temperature & $c_{Q_{1}}$ & $c_{Q_{2}}$ & $c_{H_{1}}$ & $c_{H_{2}}$ & Effective charges\tabularnewline
\hline 
\hline 
$T>T_{B}$ & 0 & 0 & $-1$ & $-\frac{2}{3}$ & $U(1)_{Y},U(1)_{E_{1}},U(1)_{E_{2}},U(1)_{E_{3}},U(1)_{B}$\tabularnewline
\hline 
$T_{\tau}<T<T_{B}$ & $-\frac{3}{4}$ & 0 & $-1$ & $-\frac{2}{3}$ & $U(1)_{Y},U(1)_{E_{1}},U(1)_{E_{2}},U(1)_{E_{3}}$\tabularnewline
\hline 
$T_{\mu}<T<T_{\tau}$ & $-\frac{8}{11}$ & $\frac{1}{33}$ & $-\frac{10}{11}$ & $-\frac{4}{11}$ & $U(1)_{Y},U(1)_{E_{1}},U(1)_{E_{2}}$\tabularnewline
\hline 
$T_{e}<T<T_{\mu}$ & $-\frac{5}{7}$ & $\frac{2}{21}$ & $-\frac{6}{7}$ & $-\frac{10}{21}$ & $U(1)_{Y},U(1)_{E_{1}}$\tabularnewline
\hline 
$T<T_{e}$ & $-\frac{12}{17}$ & $\frac{2}{17}$ & $-\frac{14}{17}$ & $-\frac{22}{51}$ & $U(1)_{Y}$\tabularnewline
\hline 
\end{tabular}

\caption{Here we list the temperature regime and the corresponding coefficients
and the effective charges that are set to zero.}
\label{tab:effective_BFC}

\end{table}

\section{Application to Leptogenesis}\label{sec:leptogenesis}

In ref.~\cite{Fong:2021xmi}, leptogenesis was studied using an effective lepton-flavor-covariant (effL) formalism where it is not necessary to follow the evolutions of quark charges. Instead, the quark interactions are captured by a temperature-dependent coefficient defined by
\begin{eqnarray}
	c_{H}(T) & \equiv & -\frac{Y_{\Delta H}}{\displaystyle \frac{1}{3}{\rm Tr}\left(Y_{\Delta Q}+Y_{\Delta U}+Y_{\Delta D}\right)-{\rm Tr}\left(Y_{\Delta\ell}+2Y_{\Delta E}\right) + 2\sum_i q^Y_{\phi_i} Y_{\Delta \phi_i} }.\label{eq:cH_def}
\end{eqnarray}
By imposing chemical equilibrium condition when a specific quark Yukawa interaction gets into equilibrium at $T_X$, ref.~\cite{Fong:2021xmi} obtained
\begin{eqnarray}
	c^{\rm eff}_{H}\left(T\right) & = &  \left(\frac{2}{3}+\frac{1}{3}e^{-T_{t}/T}\right)-\left(\frac{2}{3}-\frac{14}{23}\right)\left(1-e^{-T_{u}/T}\right)-\left(\frac{14}{23}-\frac{2}{5}\right)\left(1-e^{-T_{u-b}/T}\right)\nonumber \\
	&  & -\left(\frac{2}{5}-\frac{4}{13}\right)\left(1-e^{-T_{u-c}/T}\right)-\left(\frac{4}{13}-\frac{3}{10}\right)\left(1-e^{-T_{B_{3}-B_{2}}/T}\right)\nonumber \\
	&  & -\left(\frac{3}{10}-\frac{1}{4}\right)\left(1-e^{-T_{u-s}/T}\right)-\left(\frac{1}{4}-\frac{2}{11}\right)\left(1-e^{-T_{u-d}/T}\right),
	\label{eq:cH_eff}
\end{eqnarray}
where $T_t = 10^{15}$ GeV, $T_u = 2\times 10^{13}$ GeV, $T_{u-b} = 3\times 10^{11}$ GeV, $T_{u-c} = 2\times 10^{10}$ GeV, $T_{B_3-B_2} = 9\times 10^8$ GeV, $T_{u-s} = 3\times 10^8$ GeV and $T_{u-d} = 2\times 10^6$ GeV.
We will compare the exact result of eq.~\eqref{eq:cH_def} from solving eqs.~\eqref{eq:BEQ}--\eqref{eq:BEE} with the effective coefficient of eq.~\eqref{eq:cH_eff}. 

Finally, we will also compare the result obtained using the complete formalism against the result obtained using the effL formalism of ref.~\cite{Fong:2021xmi} from solving the Boltzmann equations for lepton charges, which are collected in Appendix~\ref{app:lepton_flavor_charges}.

\subsection{Type-I leptogenesis}

In the type-I seesaw model, a few right-handed neutrinos
$N_{i}$ is augmented to the SM as follows
\begin{eqnarray}
	-{\cal L} & \supset & \frac{1}{2}M_{i}\overline{N_{i}}N_{i}^{c}+y_{i\alpha}\overline{N_{i}}\ell_{\alpha}\epsilon H+{\rm H.c.},\label{eq:type_I}
\end{eqnarray}
where we have assumed $N_i$ to be well-seperated states i.e. the differences in their masses are much bigger than their decay widths $|M_i - M_j| \gg \Gamma_{N_{i}}, \Gamma_{N_{j}}$ such that we can work in the mass basis of $N_i$. 
As an example, we will consider $i=1,2,3$ which is motivated by SO(10) grand unified theories (GUTs).

The evolution of $Y_{N_i}$ is described by
\begin{equation}
	s{\cal H}z\frac{dY_{N_{i}}}{dz}=-\gamma_{N_{i}}\left(\frac{Y_{N_{i}}}{Y_{N_{i}}^{{\rm eq}}}-1\right), \label{eq:BEN}
\end{equation}
where $Y_{N_{i}}^{{\rm eq}}=\frac{45}{2\pi^{4}g_{\star}}\frac{M_{i}^{2}}{T^{2}}{\cal K}_{2}\left(\frac{M_{i}}{T}\right)$
with ${\cal K}_{n}\left(x\right)$ the modified Bessel function of
the second kind of order $n$ and the decay reaction density $\gamma_{N_{i}}$
is given by
\begin{eqnarray}
	\gamma_{N_{i}} & = & sY_{N_{i}}^{{\rm eq}}\Gamma_{N_{i}}\frac{{\cal K}_{1}\left(M_{i}/T\right)}{{\cal K}_{2}\left(M_{i}/T\right)},
\end{eqnarray}
with $\Gamma_{N_{i}}=\frac{\left(yy^{\dagger}\right)_{ii}M_{i}}{8\pi}$
the total decay width of $N_{i}$.~

Type-I leptogenesis can proceed through the CP-violating decays $N_i \to \ell_\alpha H$ and $N_i \to \overline\ell_\alpha H^*$. 
This is taken into account by introducing to the
right-hand side of eq.~\eqref{eq:BEell} a source and washout terms respectively given by~\cite{Blanchet:2011xq}
\begin{eqnarray}
	S^{{\rm I}} & \equiv &\sum_{i}\epsilon_{i}\gamma_{N_{i}}\left(\frac{Y_{N_{i}}}{Y_{N_{i}}^{{\rm eq}}}-1\right), \label{eq:source_typeI}\\
	W^{{\rm I}} & \equiv & - \frac{1}{2}\sum_{i}\frac{\gamma_{N_{i}}}{Y^{{\rm nor}}}\left(\frac{1}{2}\left\{ P_{i},\frac{Y_{\Delta\ell}}{g_{\ell}\zeta_{\ell}}\right\} +P_{i}\frac{Y_{\Delta H}}{g_{H}\zeta_{H}}\right),\label{eq:washout_typeI}
\end{eqnarray}
where the matrices of CP violation parameter $\epsilon_{i}$ and flavor projection $P_{i}$ are
\begin{eqnarray}
	\left(\epsilon_{i}\right)_{\alpha\beta} & = & \frac{1}{16\pi}\frac{i}{\left(yy^{\dagger}\right)_{ii}}\sum_{j\neq i}\left[\left(yy^{\dagger}\right)_{ji}y_{j\beta}y_{i\alpha}^{*}-\left(yy^{\dagger}\right)_{ij}y_{i\beta}y_{j\alpha}^{*}\right]g\left(\frac{M_{j}^{2}}{M_{i}^{2}}\right)\nonumber \\
	&  & +\frac{1}{16\pi}\frac{i}{\left(yy^{\dagger}\right)_{ii}}\sum_{j\neq i}\left[\left(yy^{\dagger}\right)_{ij}y_{j\beta}y_{i\alpha}^{*}-\left(yy^{\dagger}\right)_{ji}y_{i\beta}y_{j\alpha}^{*}\right]\frac{M_{i}^{2}}{M_{i}^{2}-M_{j}^{2}},\label{eq:CP_typeI}\\
	P_{i} & = & \frac{1}{\left(yy^{\dagger}\right)_{ii}}\left(\begin{array}{ccc}
		\left|y_{ie}\right|^{2} & y_{ie}^{*}y_{i\mu} & y_{ie}^{*}y_{i\tau}\\
		y_{ie}y_{i\mu}^{*} & \left|y_{i\mu}\right|^{2} & y_{i\mu}^{*}y_{i\tau}\\
		y_{ie}y_{i\tau}^{*} & y_{i\mu}y_{i\tau}^{*} & \left|y_{i\tau}\right|^{2}
	\end{array}\right),
\end{eqnarray}
with 
\begin{equation}
	g(x) \equiv \sqrt{x}\left[\frac{1}{1-x}+1-(1+x)\ln\frac{1+x}{x}\right].
\end{equation}

Under a change of flavor basis $\ell \to V_\ell \ell$, $y_E \to y_E V_\ell^\dagger$, $y \to y V_\ell^\dagger$, we have
\begin{eqnarray}
	\epsilon_{i} & \to & V_\ell \epsilon_{i}V_\ell^{\dagger},\;\;\;\;\;P_{i}\to V_\ell P_{i}V_\ell^{\dagger},
\end{eqnarray}
and together with eq.~\eqref{eq:flavor_transformation},
\begin{eqnarray}
	\left(Y_{\Delta \ell}\right)_{\alpha\beta} \to V_\ell \left(Y_{\Delta \ell}\right)_{\alpha\beta} V_\ell^\dagger,
    \label{eq:YDeltaell_transformation}
\end{eqnarray}
eq.~\eqref{eq:BEell} remains flavor covariant as required.

To show a numerical example, we will set $M_{\rm ref} = 10^{12}$ GeV and use the benchmark point of a minimal SO(10) GUT model with spontaneous CP violation of ref.~\cite{Babu:2025wop} where
\begin{equation}
    \left(M_1, M_2, M_3\right)= \left( 9.0\times 10^4, 1.6\times 10^{12}, 4.1\times 10^{14} \right)\; \mathrm{GeV},
\end{equation}
and the couplings $y$ and $y_E$ at scale $\mu = M_2$ and $\mu = M_1$ are given in Appendix B of ref.~\cite{Babu:2025wop}. 
Due to the RGE running between widely separated $M_2$ and $M_1$ scales, two sets of Yukawa couplings are necessary to describe leptogenesis accurately: asymmetry generation proceeds dominantly through the $N_2$ dynamics while further washout is due to the $N_1$ dynamics.

We will consider only the case where the initial $N_i$ abundance is zero.\footnote{We have verified that our final baryon asymmetry is larger by about 13\% from ref.~\cite{Babu:2024ahk} due to a mistake in the latter where the RGE running of $y$ and $y_E$ to $M_1$ scale was not actually taken into account. Incidentally, this allows us to quantify the effect of running of $y$.} 
For comparison, we will also solve the system with the effL formalism introduced in ref.~\cite{Fong:2021xmi} (see Appendix~\ref{app:lepton_flavor_charges}) where the corresponding $B-L$ charge will be denoted by $Y_{B-L}^{\rm eff}$.

\begin{figure}[ht]
    \centering
    \includegraphics[width=0.51\linewidth]{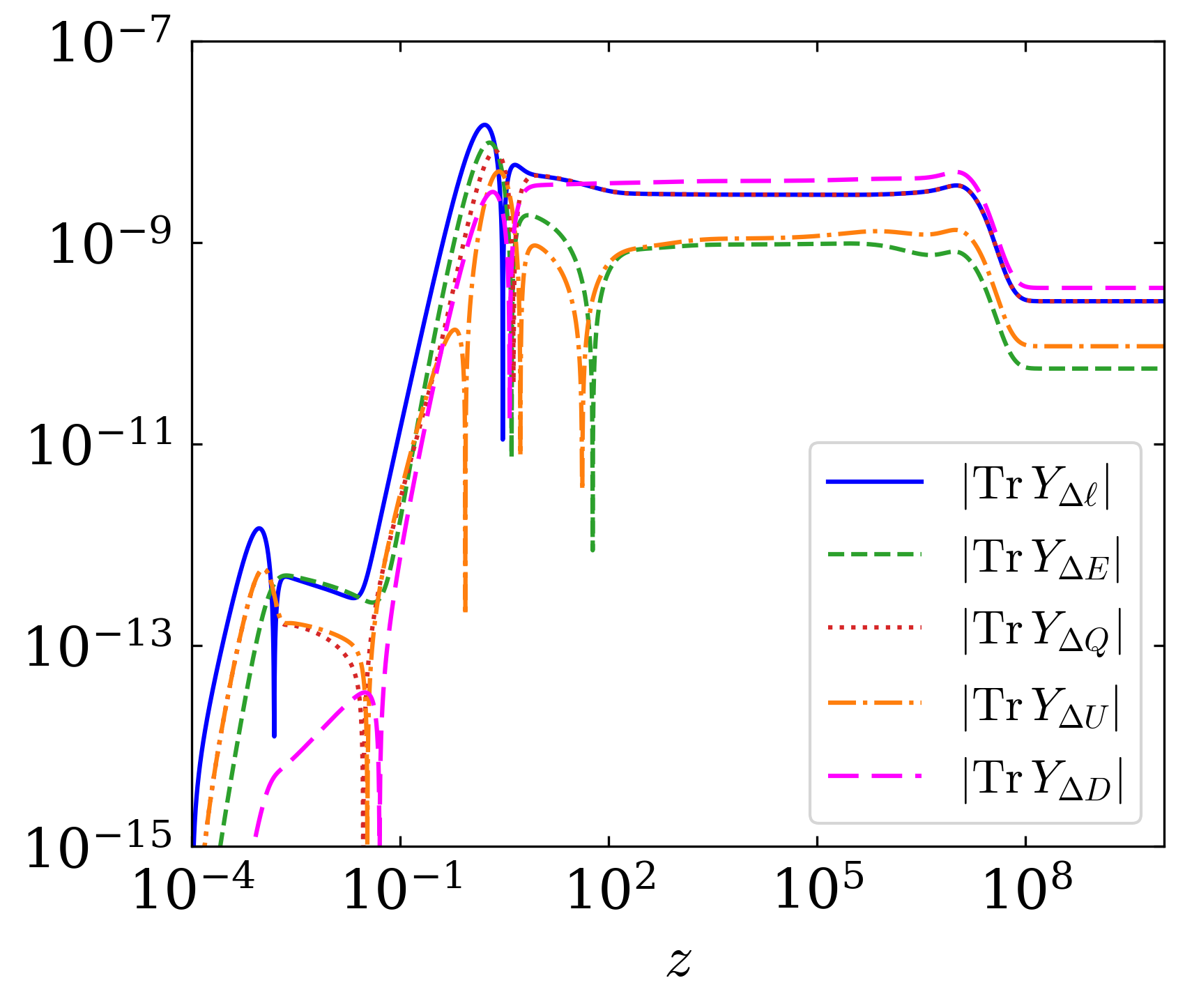}
    \includegraphics[width=0.47\linewidth]{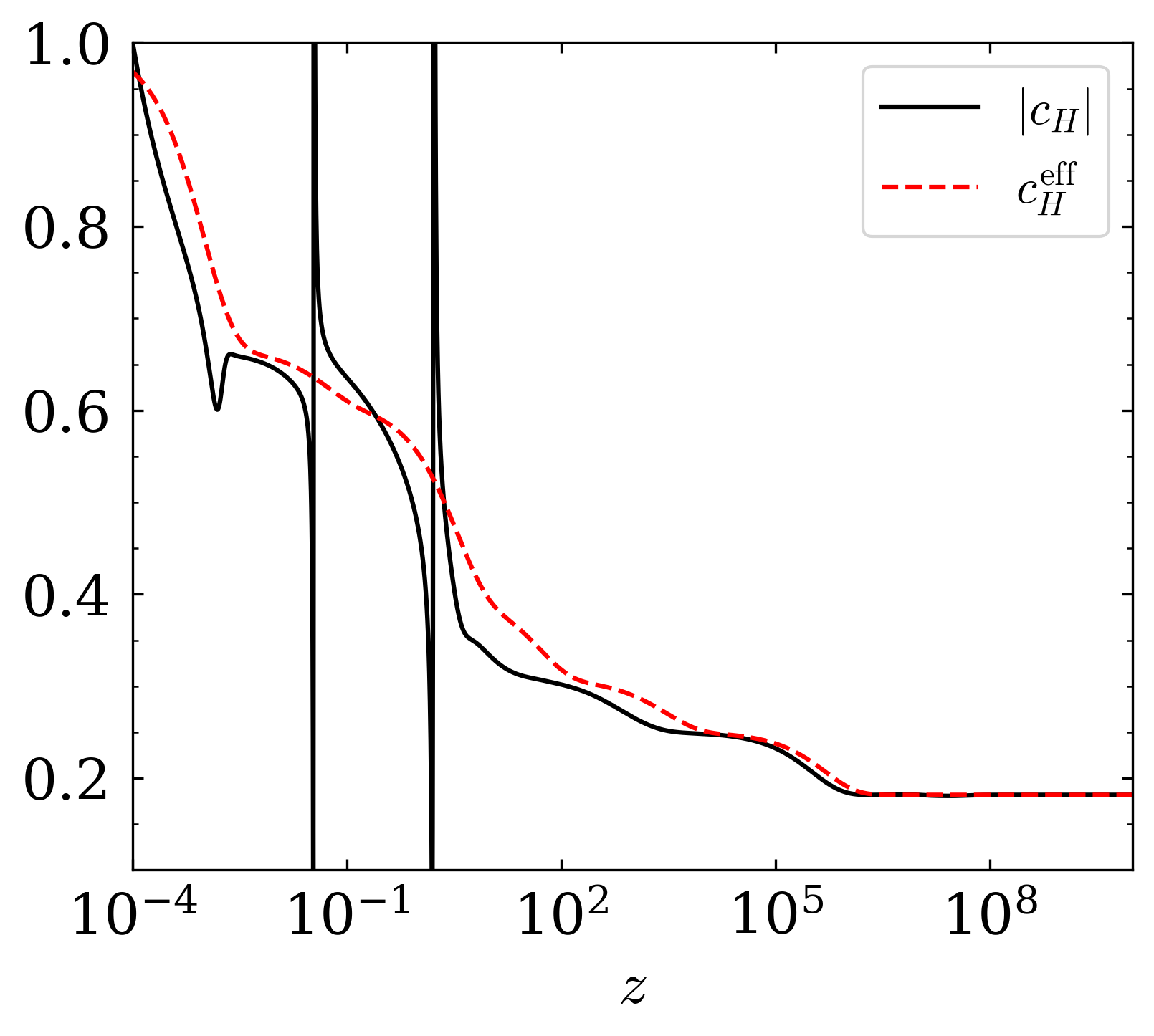}

    \caption{Left: the evolutions of the five types of SM fermion charges ${\rm Tr}\,Y_{\Delta i}$. Right: the differences between $c_H$ (solid black curve) and $c_H^{\rm eff}$ (red dashed curve) indicate deviations of quarks from chemical equilibrium.
    }
    \label{fig:typeI_charges}
\end{figure}
\begin{figure}
    \centering
    \includegraphics[width=0.6\linewidth]{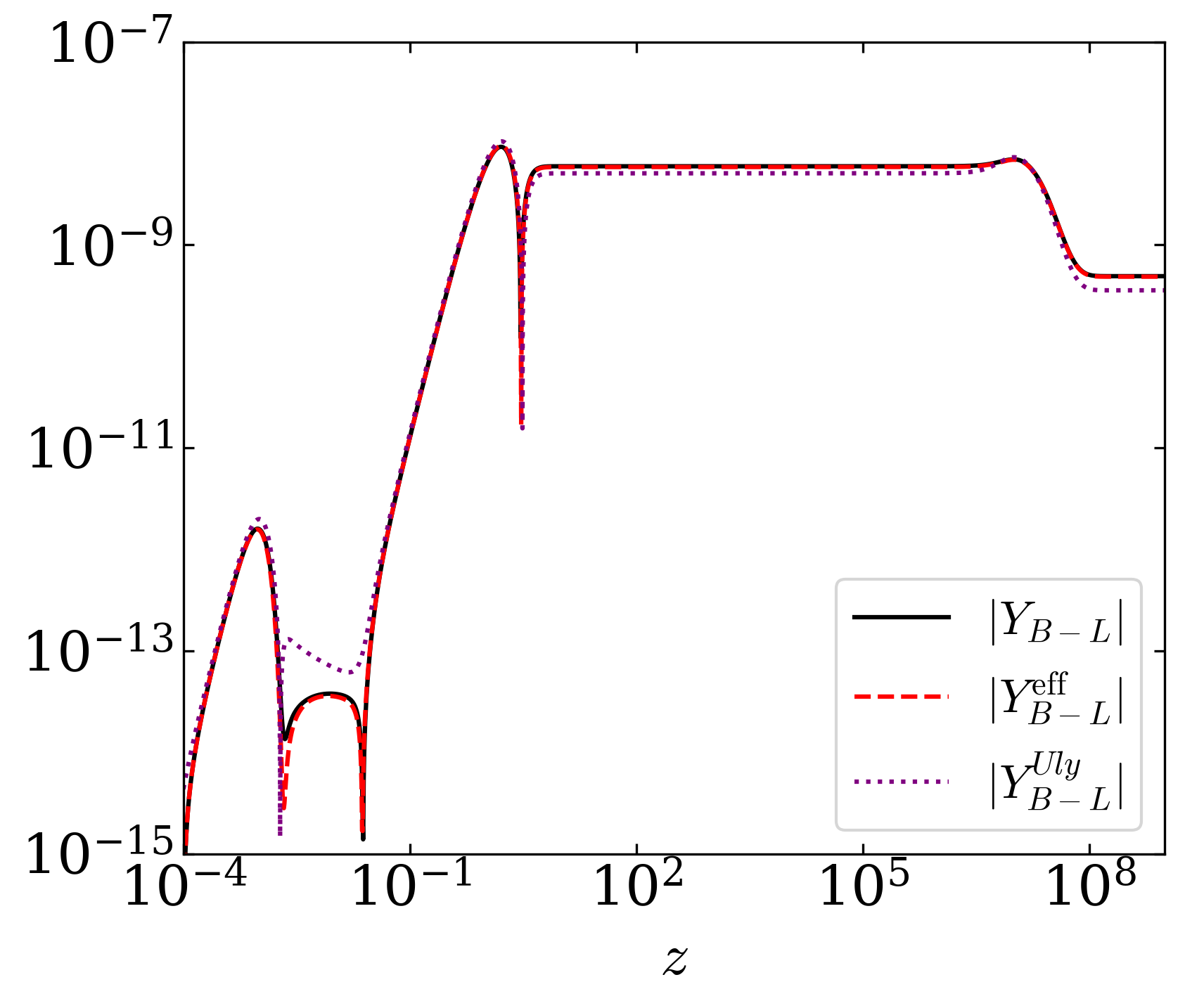}
        \caption{
        The evolution of $B-L$ charges using the complete formalism (black solid curve) and the effL formalism (red dashed curve). The final baryon asymmetry from the former (latter) is $Y ^{\rm final}_B=1.55\times 10^{-10} \,(1.53\times 10^{-10})$.
        We also compare with the result from type-I leptogenesis code ULYSSES~\cite{Granelli:2020pim} (the purple dotted line) which gives $Y ^{\rm final}_B = 1.12 \times 10^{-10}$ and the differences with our results can be attributed to spectator effects from the quark sector. 
        }
    \label{fig:BL_SSI}
\end{figure}

In the left plot of Figure~\ref{fig:typeI_charges}, we show the evolutions of the five types of SM charges ${\rm Tr}\,Y_{\Delta i}$. One observes that $|{\rm Tr}\,Y_{\Delta \ell}|$ dominates during the initial stage of leptogenesis. Although the EW sphaleron interaction gets into equilibrium at $T \lesssim 2 \times 10^{12}$ GeV ($z \lesssim 0.5$), $|{\rm Tr}\,Y_{\Delta Q}|$ and $|{\rm Tr}\,Y_{\Delta\ell}|$ start to coincide only at much lower temperature $z \gtrsim 10$ upon the completion of $N_2$-leptogenesis.
On the right plot, we show the evolution of $c_H$ given by eq.~\eqref{eq:cH_def} the complete formalism, compared to $c_H^{\rm eff}$ given by eq.~\eqref{eq:cH_eff} obtained from the effL formalism. 
The differences between them indicate deviations of quarks from chemical equilibrium while the various peaks and troughs are due to sign change in the denominator of eq.~\eqref{eq:cH_def}.

As can be observed in Figure~\ref{fig:BL_SSI}, the evolution of the $B-L$ charge in the complete formalism (black solid curve) agrees very well with the one in the effL formalism (red dashed curve).
The small differences can be attributed to the mild deviations of quarks from chemical equilibrium as shown in the right plot of Figure~\ref{fig:typeI_charges}. Those deviations translate into a difference of $\sim 1\%$
in the final baryon asymmetries: $Y ^{\rm final}_B=1.55\times 10^{-10}\,(1.53\times 10^{-10})$ for the complete (effL) formalism.
This demonstrates both the consistency of the complete formalism, and the reliability of the effL formalism in describing leptogenesis. 
Finally, we also compare our result with the one obtained from type-I leptogenesis code ULYSSES~\cite{Granelli:2020pim}, $Y_{B-L}^{Uly}$ shown as the purple dotted line in Figure~\ref{fig:BL_SSI}, with $Y ^{\rm final}_B = 1.12 \times 10^{-10}$. This value is smaller by about 28\% from our full computation and can be attributed to spectator effects of Higgs and quarks~\cite{Buchmuller:2001sr,Nardi:2005hs}.

\subsection{Type-II leptogenesis}

In the type-II seesaw model, a massive $SU(2)_L$ triplet
scalar ${\cal T}$ with hypercharge $q_{{\cal T}}^{Y}=1$ is introduced to the SM as follows
\begin{eqnarray}
	-{\cal L} & \supset & M_{{\cal T}}^{2}{\rm Tr}\left({\cal T}^{\dagger}{\cal T}\right)+\frac{1}{2}\left(f_{\alpha\beta}\overline{\ell_{\alpha}^{c}}\epsilon{\cal T}\ell_{\beta}+\mu H^{T}\epsilon{\cal T}^{\dagger}H+{\rm H.c.}\right),\label{eq:type_II}
\end{eqnarray}
where $f$ is a complex symmetric coupling and
\begin{eqnarray}
	{\cal T} & = & \left(\begin{array}{cc}
		\frac{1}{\sqrt{2}}{\cal T}^{+} & {\cal T}^{++}\\
		{\cal T}^{0} & -\frac{1}{\sqrt{2}}{\cal T}^{+}
	\end{array}\right).
\end{eqnarray}
Since ${\cal T}$ couples to two lepton doublets which in general do not align in flavor space, one needs to describe them with density matrix as first pointed out in ref. \cite{Lavignac:2015gpa}. 

The evolutions of $Y_{\Sigma{\cal T}}\equiv Y_{{\cal T}}+Y_{{\cal T}^{\dagger}}$
and $Y_{\Delta{\cal T}}\equiv Y_{{\cal T}}-Y_{{\cal T}^{\dagger}}$
are described by~\cite{Lavignac:2015gpa}
\begin{eqnarray}
	s{\cal H}z\frac{dY_{\Sigma{\cal T}}}{dz} & = & -\gamma_{D}\left(\frac{Y_{\Sigma{\cal T}}}{Y_{\Sigma{\cal T}}^{{\rm eq}}}-1\right)-2\gamma_{A}\left(\frac{Y_{\Sigma{\cal T}}^{2}}{Y_{\Sigma{\cal T}}^{{\rm eq},2}}-1\right),
	\label{eq:BE_sumT}\\
	s{\cal H}z\frac{dY_{\Delta{\cal T}}}{dz} & = & -\gamma_{D}\left(\frac{Y_{\Delta{\cal T}}}{Y_{\Sigma{\cal T}}^{{\rm eq}}}+B_{\ell}\frac{{\rm Tr}\left(ff^{\dagger}Y_{\Delta\ell}\right)}{{\rm Tr}\left(ff^{\dagger}\right)Y^{{\rm nor}}g_{\ell}\zeta_{\ell}}-B_{H}\frac{Y_{\Delta H}}{Y^{{\rm nor}}g_{H}\zeta_{H}}\right),\label{eq:BE_DeltaT}
\end{eqnarray}
where eq.~\eqref{eq:YH} becomes
\begin{eqnarray}
	Y_{\Delta H} & = & 
	\frac{1}{3}{\rm Tr}\left[-Y_{\Delta Q}-4Y_{\Delta U}+2Y_{\Delta D}\right]
	+{\rm Tr}\left[Y_{\Delta\ell}+2Y_{\Delta E}\right]
	- 2 Y_{\Delta {\cal T}}.
	\label{eq:YH_T}
\end{eqnarray}
The branching ratios for
the decays of ${\cal T}$ to lepton doublets ${\cal T}^{\dagger}\to\ell_{\alpha}\ell_{\beta}$
and Higgses ${\cal T}\to HH$ are respectively
\begin{equation}
	B_{\ell} = \frac{{\rm Tr}\left(ff^{\dagger}\right)}{{\rm Tr}\left(ff^{\dagger}\right)+\frac{\left|\mu\right|^{2}}{M_{{\cal T}}^{2}}},\quad
	B_{H} = \frac{\frac{\left|\mu\right|^{2}}{M_{{\cal T}}^{2}}}{{\rm Tr}\left(ff^{\dagger}\right)+\frac{\left|\mu\right|^{2}}{M_{{\cal T}}^{2}}},
\end{equation}
while $Y_{\Sigma{\cal T}}^{{\rm eq}}=Y_{{\cal T}}^{{\rm eq}}+Y_{{\cal T}^{\dagger}}^{{\rm eq}}=\frac{135}{2\pi^{4}g_{\star}}z^{2}{\cal K}_{2}\left(z\right)$
and the decay reaction density $\gamma_{D}$ is given by
\begin{eqnarray}
	\gamma_{D} & = & sY_{\Sigma {\cal T}}^{{\rm eq}}\Gamma_{{\cal T}}\frac{{\cal K}_{1}\left(z\right)}{{\cal K}_{2}\left(z\right)},
\end{eqnarray}
with the total decay width of $\cal T$ given by
\begin{eqnarray}
	\Gamma_{{\cal T}} & = & \frac{M_{{\cal T}}}{32\pi}\left[{\rm Tr}\left(ff^{\dagger}\right)+\frac{\left|\mu\right|^{2}}{M_{{\cal T}}^{2}}\right].
\end{eqnarray}

Since $\cal T$ experiences $SU(2)_L \times U(1)_Y$ gauge interactions,  we need to take into account the scatterings ${\cal T}{\cal T}^\dagger \leftrightarrow \psi\bar\psi$ where $\psi$ refers to the SM fields.
The corresponding reaction density in eq.~\eqref{eq:BE_sumT} is 
\begin{eqnarray}
	\gamma_{A} & = & \frac{M_{{\cal T}}^{4}}{64\pi^{4}z}\int_{4}^{\infty}dx\sqrt{x}{\cal K}_{1}\left(z\sqrt{x}\right)\hat{\sigma}_{A}\left(x\right),
\end{eqnarray}
where the reduced cross section is given by \cite{Hambye:2005tk}
\begin{eqnarray}
	\hat{\sigma}_{A}\left(x\right) & = & \frac{1}{16\pi x^{2}}\bigg\{ \sqrt{x}\sqrt{x-4}\left[96g_{2}^{2}g_{Y}^{2}\left(x+4\right)+g_{Y}^{4}\left(65x-68\right)+2g_{2}^{4}\left(172+65x\right)\right] \nonumber \\
	&  & \!\!\!\! \left.-96\left[4g_{2}^{2}g_{Y}^{2}\left(x-2\right)+g_{Y}^{4}\left(x-2\right)+4g_{2}^{4}\left(x-1\right)\right]\ln\left(\frac{\sqrt{x-4}\sqrt{x}+x}{2}-1\right)\right\} .
\end{eqnarray}

In order to realize leptogenesis through the decays of ${\cal T}^{\dagger}\to\ell_{\alpha}\ell_{\beta}$
and ${\cal T}\to HH$, one loop correction is needed. This can be achieved by considering a Weinberg operator generated at $\Lambda \gg M_{\cal T}$
\begin{eqnarray}
	{\cal L} & \supset & \frac{1}{4}\frac{\kappa_{\alpha\beta}}{\Lambda}\overline{\ell_{\alpha}^{c}}\epsilon H\,H^{T} \epsilon \ell_{\beta}+{\rm H.c.},
\end{eqnarray}
where $\kappa$ is a symmetric complex coupling.
After the EW symmetry breaking where the Higgs acquire a $v = \langle H\rangle = 174$ GeV, there are two contributions to the light neutrino mass matrix, from integrating out the scalar triplet $\cal T$ and from the Weinberg operator:
\begin{eqnarray}
	m_{\nu} & = & m_{{\cal T}}+m_{\Lambda},
\end{eqnarray}
where 
\begin{equation}
	m_{{\cal T}} \equiv \frac{1}{2}\mu f\frac{v^{2}}{M_{\cal T}^{2}},\quad
	m_{\Lambda} \equiv \frac{1}{2}\kappa\frac{v^{2}}{\Lambda}.
\end{equation}

The source and washout terms to be introduced to the right-hand side of eq.~\eqref{eq:BEell} are respectively given by \cite{Lavignac:2015gpa}
\begin{eqnarray}
	S^{{\rm II}} & \equiv & \epsilon\gamma_{D}\left(\frac{Y_{\Sigma{\cal T}}}{Y_{{\cal T}}^{{\rm eq}}}-1\right),\label{eq:source_typeII}\\
	W^{{\rm II}} & \equiv & - \frac{2\gamma_{D}}{{\rm Tr}\left(ff^{\dagger}\right)+\frac{\left|\mu\right|^{2}}{M_{{\cal T}}^{2}}}
	\left[\left(f^\dagger f \right)\frac{Y_{\Delta{\cal T}}}{Y_{{\cal T}}^{{\rm eq}}}
	+\frac{1}{4Y^{{\rm nor}}g_{\ell}\zeta_{\ell}}\left(2f^\dagger Y_{\Delta\ell}^T f + Y_{\Delta\ell} f^\dagger f + f^\dagger f Y_{\Delta\ell} \right)\right],\label{eq:washout_typeII}
\end{eqnarray}
where the matrix of CP violation parameter is 
\begin{eqnarray}
	\epsilon & = & \frac{i}{8\pi}\frac{M_{\cal T}}{v^{2}}\sqrt{B_{\ell}B_{H}}\frac{m_{\Lambda}^{\dagger} m_{{\cal T}} - m_{{\cal T}}^\dagger m_{\Lambda}}{\sqrt{{\rm Tr}\left(m_{{\cal T}}^{\dagger}m_{{\cal T}}\right)}}.\label{eq:CP_typeII}
\end{eqnarray}
Under a change of flavor basis $\ell \to V_\ell \ell$, $y_E \to y_E V_\ell^\dagger$, $y \to y V_\ell^\dagger$, $ f \to V_\ell^* f V_\ell^\dagger $, $\kappa \to V_\ell^* \kappa V_\ell^\dagger$, we have
\begin{equation}
	f^{\dagger}f \to V_\ell f^{\dagger} f V_\ell^{\dagger},\quad \epsilon \to V_\ell \epsilon V_\ell^\dagger,
\end{equation}
and together with eq.~\eqref{eq:YDeltaell_transformation}, eq.~\eqref{eq:BEell} remains flavor covariant as expected.

For a numerical example, we set
\begin{eqnarray}
	m_{{\cal T}} & = & i m_{\nu}\implies m_{\Lambda}=\left(1-i\right)m_{\nu},\\
	M_{\cal T} & = & 10^{11}\,{\rm GeV} = M_{\rm ref},\quad \mu = 0.1M_{{\cal T}},
\end{eqnarray}
where the neutrino mass matrix is taken to be
\begin{eqnarray}
	m_{\nu} & = & r\, U_{{\rm PMNS}}^{*}{\rm diag}\left(m_{1},m_{2},m_{3}\right)U_{{\rm PMNS}}^{\dagger},
\end{eqnarray}
with $r=1.3$ (to take into account the RGE evolution), $m_{1}=10^{-3}$ eV, $m_2^2-m_1^2=7.5\times 10^{-5}$ eV$^2$, $m_3^2 - m_1^2 = 2.5 \times 10^{-3}$ eV$^2$ and the rest of the parameters in the lepton mixing matrix $U_{\rm PMNS}$ are taken from the normal mass ordering global fit of ref.~\cite{Esteban:2024eli}.
The charged lepton Yukawa is taken to be the value at $\mu = 10^{12}$ GeV~\cite{Antusch:2025fpm}: 
\begin{equation}
	y_{E}={\rm diag}\left(2.8\times10^{-6},5.9\times10^{-4},1.0\times10^{-2}\right).
\end{equation}

\begin{figure}[ht]
    \centering
        \includegraphics[width=0.50\linewidth]{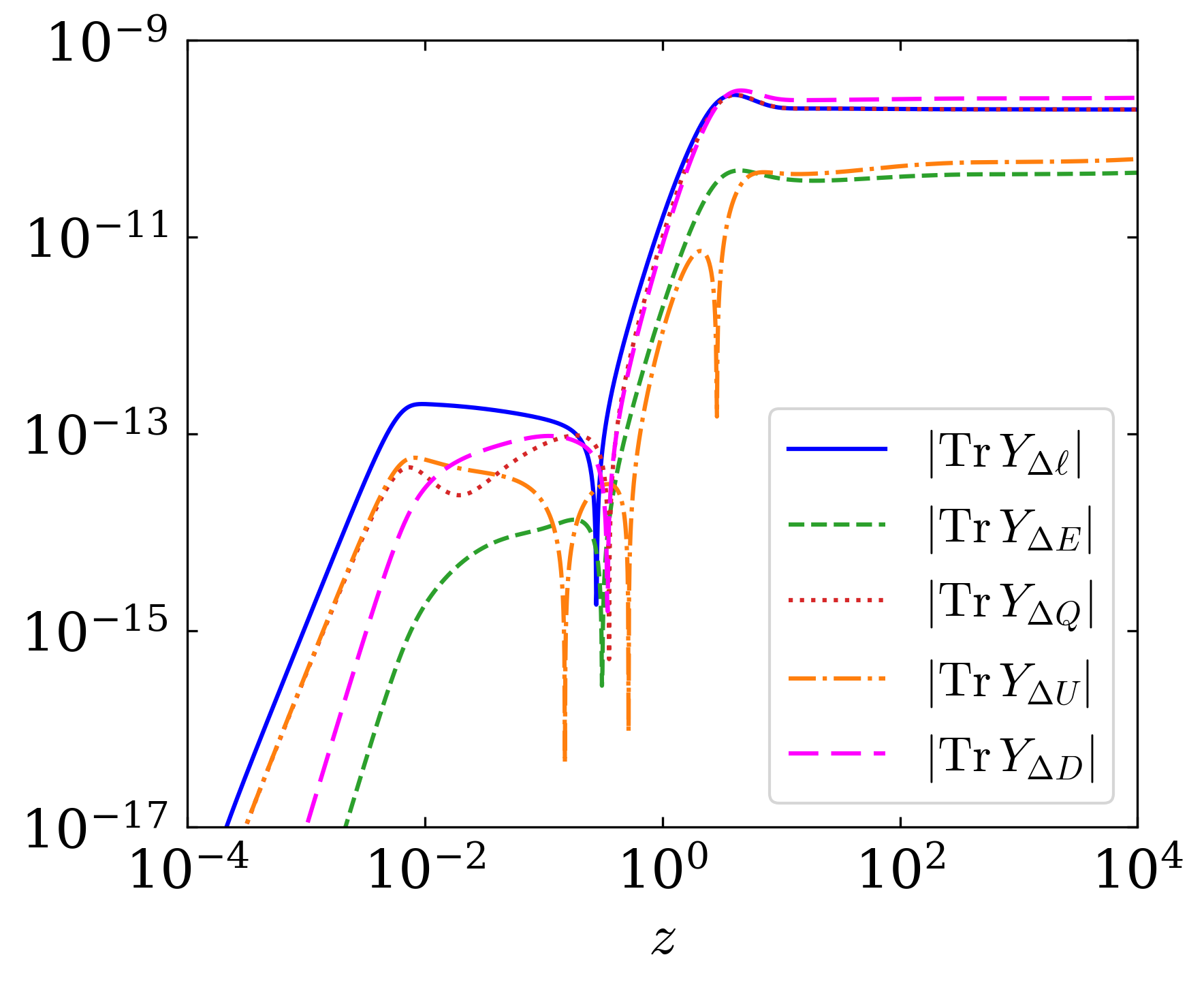}
        \includegraphics[width=0.48\linewidth]{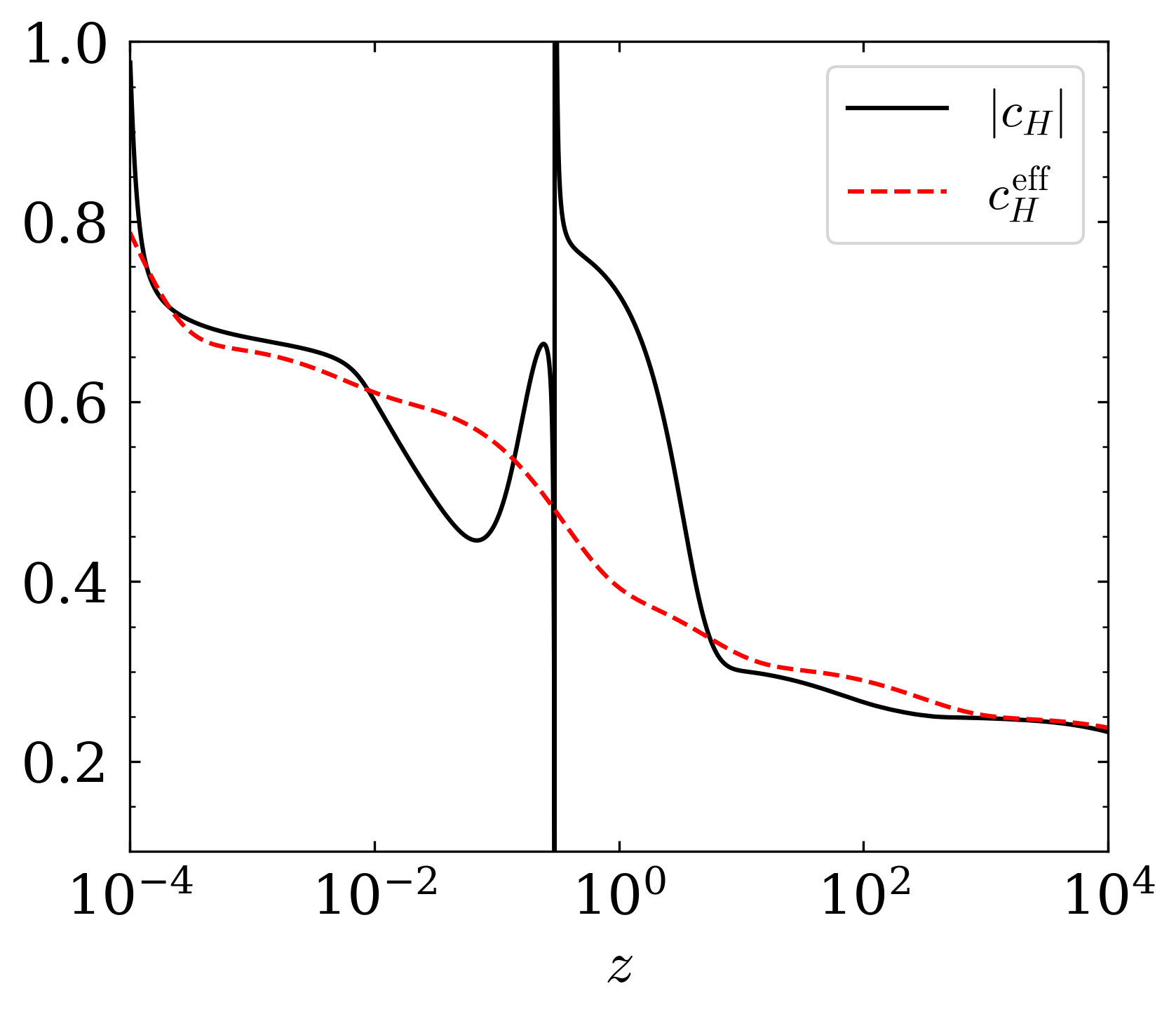}

    \caption{Left: the evolution of the five types of SM fermion charges ${\rm Tr}\,Y_{\Delta i}$. Right: the differences between $c_H$ (solid black curve) and $c_H^{\rm eff}$ (red dashed curve) indicate deviations of quarks from chemical equilibrium.}
    \label{fig:typeII_charges}
\end{figure}

\begin{figure}
    \centering    \includegraphics[width=0.6\linewidth]{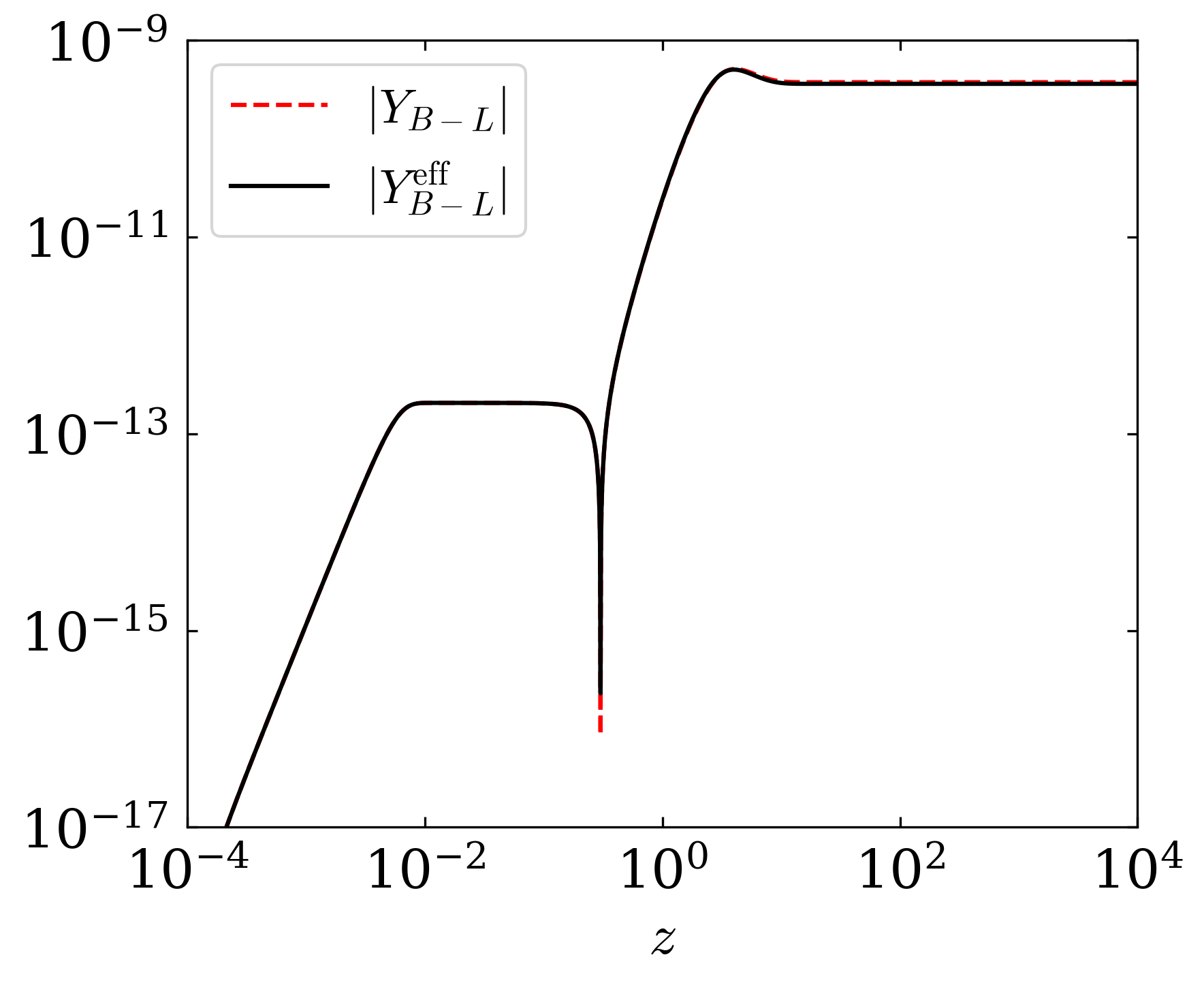}
    \caption{The evolution of $B-L$ charges using the complete formalism (black solid curve) and the effL formalism (red dashed curve). The final baryon asymmetry from the former (latter) is $Y ^{\rm final}_B=1.18\times 10^{-10} \,(1.14\times 10^{-10})$.}
    \label{fig:BL_SSII}
\end{figure}

The left plot of Figure~\ref{fig:typeII_charges} shows the evolution of the SM fermion charges ${\rm Tr}Y_{\Delta i}$.
One clear difference in comparison to the type-I leptogenesis is the suppression of $|{\rm Tr}\,Y_{\Delta \ell}|$ in the initial stage due to the inefficiency in asymmetry generation when $\cal T$ is kept closed to equilibrium by the gauge interactions.
Similarly to the type-I leptogenesis, $|{\rm Tr}\,Y_{\Delta Q}|$ and $|{\rm Tr}\,Y_{\Delta\ell}|$ start to coincide after the end of leptogenesis at $z \gtrsim 10$. 
In the right plot, the differences between $c_H$ and $c_H^{\rm eff}$ are slightly more pronounced in the early stage than in the type-I seesaw leptogenesis, indicating a larger deviations from quark chemical equilibrium.

In Figure~\ref{fig:BL_SSII}, we show the evolution of the $B-L$ charge in the complete formalism (black solid curve) and in the effL formalism (red dashed curve). 
 
The deviations from quark chemical equilibrium result in a difference of $\sim 3\%$ in the final baryon asymmetries: $Y ^{\rm final}_B=1.18\times 10^{-10} \,(1.14\times 10^{-10})$ in the complete (effL) formalism.
This again demonstrates the consistency between the two formalisms.

\section{Application to Cloistered Baryogenesis}\label{sec:cloistered}

For cloistered baryogenesis~\cite{AristizabalSierra:2013lyx}, we select the model in which a scalar $\tilde U$ carrying the same SM quantum number as $U_\alpha$ is introduced to the type-I seesaw model
\begin{eqnarray}
	-{\cal L} & \supset & \frac{1}{2}M_{i}\overline{N_{i}}N_{i}^{c}+\eta_{i\alpha}\overline{N_{i}}U^c_{\alpha}\tilde U + {\rm H.c.},\label{eq:u_cloistered}
\end{eqnarray}
where we have suppressed the neutrino Yukawa interaction.
$\tilde U$-cloistered baryogenesis can proceed through the CP-violating decays $N_i \to \overline U_\alpha \tilde U$ and $N_i \to U_\alpha \tilde U^*$.
Since the coupling $\eta$ is not connected to neutrino mass, the CP violation is not subjected to the Davidson-Ibarra bound~\cite{Davidson:2002qv} and as a result, the seesaw scale can be much lower.

For illustration, we will consider $i=2$ with $M_{\rm ref} = M_1 = 5\times 10^7$ GeV, $M_2 = 2 M_1$ and
\begin{eqnarray}
	\eta & = & 10^{-2} \left(\begin{array}{ccc}
		0.05 e^{-i \pi/2} & 0.03 e^{-i \pi/3} & 0.02 e^{-i \pi/4}\\
		8e^{-i \pi/3} & 3e^{-i \pi/4} & 2e^{-i \pi/5}
	\end{array}\right).
\end{eqnarray}
In this case, due to the Davidson-Ibarra bound, type-I leptogenesis is in general not successful (too little baryon asymmetry is produced)~\cite{Antusch:2011nz}. With the choice of $\eta$ above, due to the neutrino mass constraint, the branching ratios for $N_i \to \ell_\alpha H$ is in general subdominant compared to $N_i \to U_\alpha \tilde U^*$.
Therefore for $N_i$ decay reaction density, we will consider only the $N_i \to U_\alpha \tilde U^*$ channel
\begin{eqnarray}
	\gamma_{N_{i}}^{\tilde U} & = & sY_{N_{i}}^{{\rm eq}}\Gamma^{\tilde U}_{N_{i}}\frac{{\cal K}_{1}\left(M_{i}/T\right)}{{\cal K}_{2}\left(M_{i}/T\right)},
\end{eqnarray}
with $\Gamma^{\tilde U}_{N_{i}}=\frac{3\left(\eta \eta^{\dagger}\right)_{ii}M_{i}}{16\pi}$.

The evolution of $Y_{N_i}$ is described by eq.~\eqref{eq:BEN} but with the replacement $\gamma_{N_i} \to \gamma^{\tilde U}_{N_i}$. 
Then we add to the right-hand side of eq.~\eqref{eq:BEU} the following source and washout terms
\begin{eqnarray}
	S^{{\rm \tilde U}} & \equiv & -\sum_{i}\epsilon^{\tilde U}_{i}\gamma^{\tilde U}_{N_{i}}\left(\frac{Y_{N_{i}}}{Y_{N_{i}}^{{\rm eq}}}-1\right),\\
	W^{{\rm \tilde U}} & \equiv & -\frac{1}{2}\sum_{i}\frac{\gamma^{\tilde U}_{N_{i}}}{Y^{{\rm nor}}}\left(\frac{1}{2}\left\{ P^{\tilde U}_{i},\frac{Y_{\Delta U}}{g_{U}\zeta_{U}}\right\} - P^{\tilde U}_{i}\frac{Y_{\Delta \tilde U}}{g_{\tilde U}\zeta_{\tilde U}}\right),
\end{eqnarray}
where $g_U = g_{\tilde U} = 3$, $\zeta_U = \zeta_{\tilde U}/2 = 1$, and the matrices of CP violation parameter $\epsilon^{\tilde U}_{i}$ and flavor projection $P^{\tilde U}_{i}$ are given by
\begin{eqnarray}
	\left(\epsilon^{\tilde U}_{i}\right)_{\beta\alpha} & = & \frac{1}{16\pi}\frac{i}{\left(\eta\eta^{\dagger}\right)_{ii}}\sum_{j\neq i}\left[\left(\eta\eta^{\dagger}\right)_{ji}\eta_{j\beta}\eta_{i\alpha}^{*}-\left(\eta\eta^{\dagger}\right)_{ij}\eta_{i\beta}\eta_{j\alpha}^{*}\right]h\left(\frac{M_{j}^{2}}{M_{i}^{2}}\right)\nonumber \\
	&  & +\frac{3}{32\pi}\frac{i}{\left(\eta\eta^{\dagger}\right)_{ii}}\sum_{j\neq i}\left[\left(\eta\eta^{\dagger}\right)_{ij}\eta_{j\beta}\eta_{i\alpha}^{*}-\left(\eta\eta^{\dagger}\right)_{ji}\eta_{i\beta}\eta_{j\alpha}^{*}\right]\frac{M_{i}^{2}}{M_{i}^{2}-M_{j}^{2}},\label{eq:CP_cloistered}\\
	P^{\tilde U}_{i} & = & \frac{1}{\left(\eta\eta^{\dagger}\right)_{ii}}\left(\begin{array}{ccc}
		\left|\eta_{ie}\right|^{2} & \eta_{ie}\eta_{i\mu}^* & \eta_{ie}\eta_{i\tau}^*\\
		\eta_{ie}^*\eta_{i\mu} & \left|\eta_{i\mu}\right|^{2} & \eta_{i\mu}\eta_{i\tau}^*\\
		\eta_{ie}^*y_{i\tau} & \eta_{i\mu}^ *\eta_{i\tau} & \left|\eta_{i\tau}\right|^{2}
	\end{array}\right),
\end{eqnarray}
with 
\begin{equation}
	h(x) \equiv \sqrt{x}\left[\frac{3}{2}\frac{1}{1-x}+1-(1+x)\ln\frac{1+x}{x}\right].
\end{equation}
Finally, we will also need the Boltzmann equation of $Y_{\Delta \tilde U}$ as follows
\begin{eqnarray}
	s{\cal H}z\frac{dY_{\Delta \tilde U}}{dz} & = & - {\rm Tr}\left(S^{\tilde U} + W^{\tilde U}\right),\label{eq:BEtU}
\end{eqnarray}
while eq.~\eqref{eq:YH} becomes
\begin{eqnarray}
	Y_{\Delta H} & = & 
	\frac{1}{3}{\rm Tr}\left[-Y_{\Delta Q}-4Y_{\Delta U}+2Y_{\Delta D}\right]
	+{\rm Tr}\left[Y_{\Delta\ell}+2Y_{\Delta E}\right]
	- \frac{4}{3} Y_{\Delta \tilde U}.
\end{eqnarray}

Under a change of flavor basis $U \to V_U U$, $y_U \to V_U y_U$, $\eta \to \eta V_U^T$, we have
\begin{eqnarray}
	\epsilon^{\tilde U}_{i} & \to & V_U \epsilon^{\tilde U}_{i}V_U^{\dagger},\;\;\;\;\;P^ {\tilde U}_{i}\to V_U P^{\tilde U}_{i}V_U^{\dagger},
\end{eqnarray}
and together with eq.~\eqref{eq:flavor_transformation},
\begin{eqnarray}
	\left(Y_{\Delta U}\right)_{\alpha\beta} \to V_U \left(Y_{\Delta U}\right)_{\alpha\beta} V_U^\dagger,
    \label{eq:YDeltaU_transformation}
\end{eqnarray}
eq.~\eqref{eq:BEU} remains flavor covariant as required. Since eq.~\eqref{eq:BEtU} involves the trace, it is clearly flavor-basis-invariant as $Y_{\Delta \tilde U}$ should be.

From the set of Boltzmann equations \eqref{eq:BEQ}--\eqref{eq:BEE} and \eqref{eq:BEtU}, we can verify the following conservation law
\begin{equation}
    \frac{d}{dz}\left(Y_{B-L} + \frac{1}{3}Y_{\Delta \tilde U}\right)=0,\label{eq:conservation_B-L}
\end{equation}
since by ignoring the neutrino Yukawa interaction, there is a conserved total $B-L$ charge where $\tilde U$ carries baryon number $q^B_{\tilde U} = 1/3$. Notice that in writing $Y_{B-L}$ above, we only include the SM fields. Due to nontrivial quark flavor correlation, to obtain the exact result, one will still need to solve the whole set of Boltzmann equations.\footnote{In ref.~\cite{AristizabalSierra:2013lyx}, by ignoring flavor correlation and imposing chemical equilibrium conditions, one can obtain a closed set of Boltzmann equations in terms of only $Y_{N_i}$ and $Y_{\Delta \tilde U}$.} 

One interesting constraint is that $\tilde U$ should decay back to the SM up-type quarks before the BBN since the determination of baryon asymmetry from the BBN and CMB measurements cannot differ by more than a few percents. Therefore $\tilde U$ is allowed to be long-lived on particle collider timescale~\cite{AristizabalSierra:2013lyx}. 

After the freezeout of the EW sphaleron interaction, besides eq.~\eqref{eq:YB_YB-L}, we should include the baryon number carried by $\tilde U$ such that the final baryon asymmetry is
\begin{equation}
    Y_B^{\rm final} = \left(0.315 Y_{B-L}+ \frac{1}{3}Y_{\Delta \tilde U}\right)\Big|_{T = 132\,{\rm GeV}} = 0.228 Y_{\Delta \tilde U}\Big|_{T = 132\,{\rm GeV}},
\end{equation}
where we have used the conservation law eq.~\eqref{eq:conservation_B-L} in the second equality assuming zero initial asymmetries.

\begin{figure}
    \centering
    \includegraphics[width=0.50\linewidth]{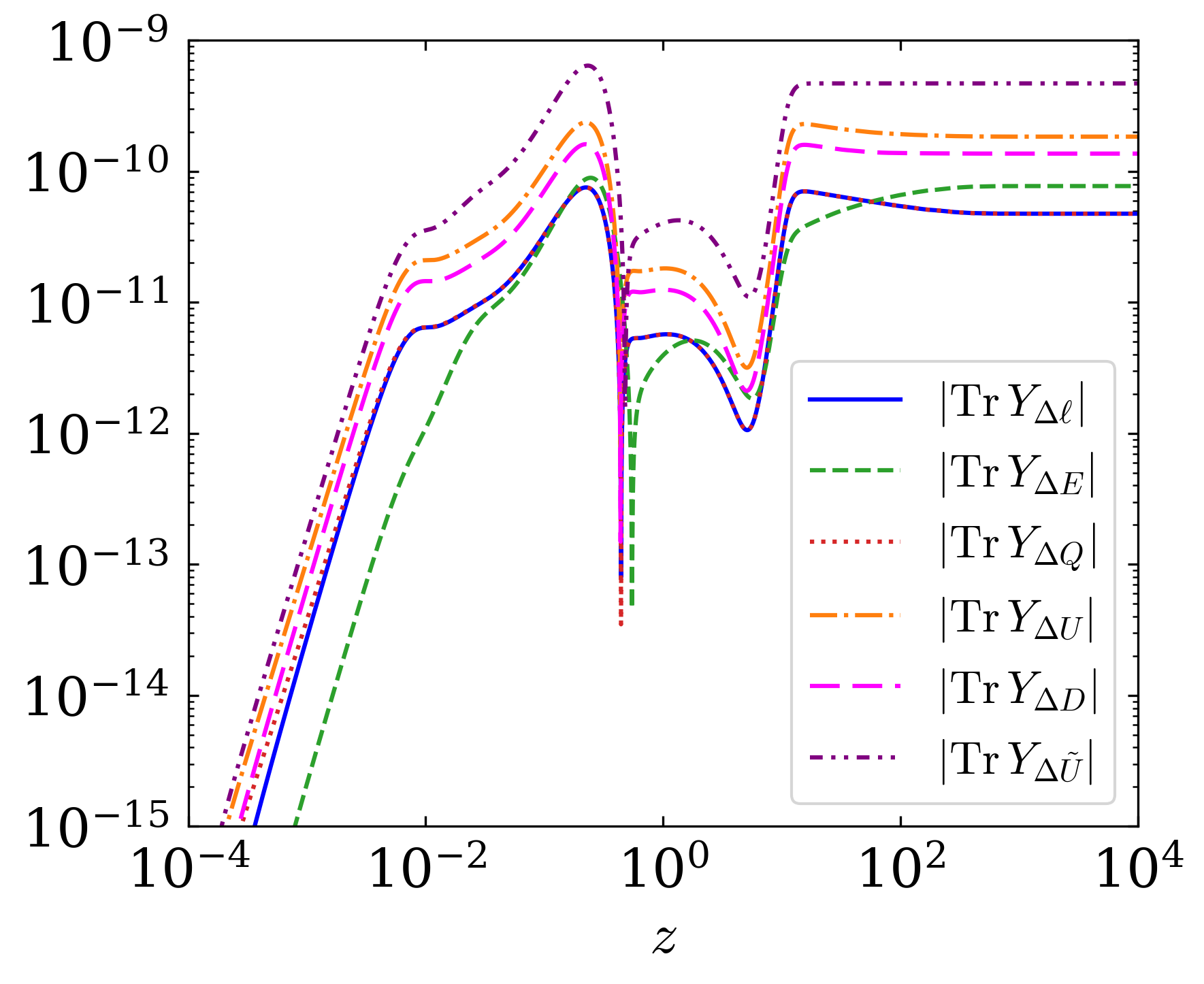}
    \includegraphics[width=0.49\linewidth]{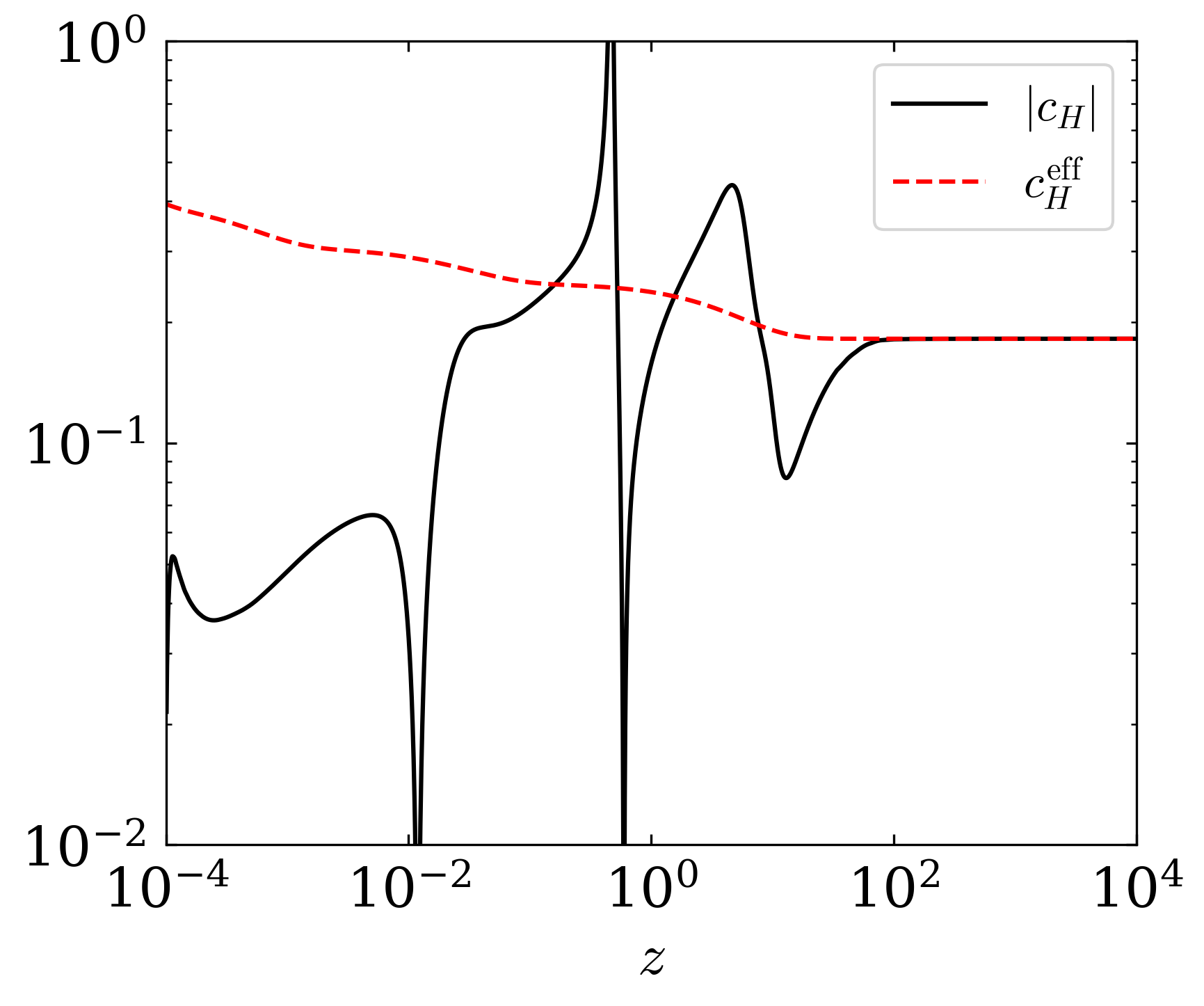}
        \caption{Left: the evolution of $|Y_{\Delta \tilde U}|$ and the five SM fermion charges $|{\rm Tr}Y_{\Delta i}|$. 
        Right: $c_H$ deviates significantly from $c_H^{\rm eff}$ during baryogenesis $z \lesssim 100$ when quarks deviate from chemical equilibrium.}
    \label{fig:cloistered}
\end{figure}

In the left plot of Figure~\ref{fig:cloistered}, we show the evolutions of $Y_{\Delta \tilde U}$ and the SM fermion charges ${\rm Tr}\,Y_{\Delta i}$. Since the asymmetry is sourced directly in $\tilde U$ and the up-type quarks, $|Y_{\Delta \tilde U}|$ and $|{\rm Tr}Y_{\Delta U}|$ remain dominant throughout the evolution. One also notice that $|{\rm Tr}Y_{\Delta \ell}|$ and $|{\rm Tr}Y_{\Delta Q}|$ almost always coincide since the EW sphaleron interaction gets into equilibrium at $T \lesssim 2 \times 10^{12}$ GeV. 
On the right plot of Figure~\ref{fig:cloistered}, we see an interesting feature when comparing  $c_H$ and $c_H^{\rm eff}$. Since $c_H^{\rm eff}$ is derived assuming chemical equilibrium of quarks, we see that $c_H$ deviates significantly from $c_H^{\rm eff}$ when quarks are out of chemical equilibrium during baryogenesis up till around $z\sim 100$. This indicates the necessity of the complete or effB formalism to accurately describe the scenario as we will discuss next.

\begin{figure}
    \centering
    \includegraphics[width=0.50\linewidth]{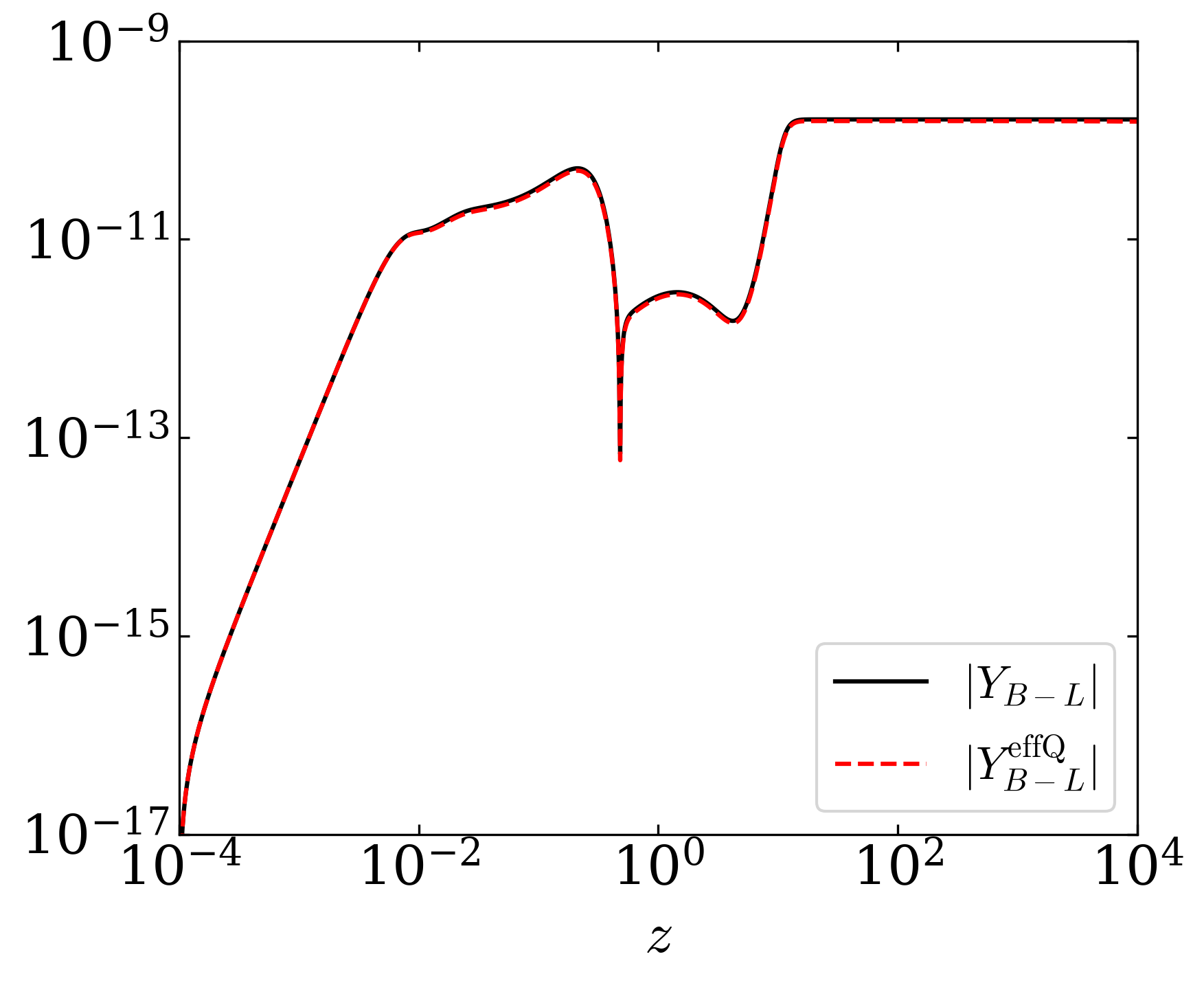}
    \caption{The evolution of $B-L$ charges using the complete formalism (black solid curve) and the effQ formalism (red dashed curve). The final baryon asymmetry from the former (latter) is $Y^{\rm final}_{B}=1.07\time10^{-10}$ ($Y^{\rm final}_B=1.03\times10^{-10}$).
    }
    \label{fig:BLClois}
\end{figure}

Since the new interactions affect only the quark sector through the up-type right handed quarks asymmetry, it is appropriate to use the effQ formalism developed in Section~\ref{sec:effQ}.
As shown in Figure~\ref{fig:BLClois}, the evolution of the $B-L$ charge obtained with the complete formalism (black solid curve) and with the effQ formalism (red dashed curve) is in very good agreement. The deviations from the complete result lead to only a $\sim4\%$ difference in the final baryon asymmetry, with $Y^{\rm final}_{B}=1.07\times10^{-10}$ ($1.03\times10^{-10}$) obtained in the complete (effQ) formalism. This supports the validity of the effQ description for the cloistered baryogenesis involving quarks.

\section{Conclusions} \label{sec:conclusions}

In this work, we have presented the complete flavor-covariant Boltzmann equations to describe the evolution of the Standard Model quark and lepton flavor charges, using the measured Standard Model parameters as inputs. 
To verify the formalism, we have applied it to the well-studied type-I and type-II leptogenesis scenarios, which have showed good agreements with the results obtained under the approximation where quarks achieve chemical equilibrium at various temperatures. 

For baryogenesis scenario in which new physics couples exclusively to the quark sector, we have also derived a new effective quark-flavor-covariant description in terms of quark charges only, obtaining a reduced system of evolution equations that reproduces the evolution of the full flavor-covariant formalism while reducing the computational cost. 
We have obtained good agreement between the two formalisms when applying them to a cloistered baryogenesis scenario where the source of asymmetry generation is in the up-type right-handed quark sector. 

To our knowledge, this work represents the first studies in leptogenesis and cloistered baryogenesis where all the relevant Standard Model interactions have been consistently taken into account.
Finally, we have released the first generic baryogenesis code \texttt{BOLEH} that can be readily applied to other models beyond the ones presented here.

\section*{Acknowledgments}
The authors acknowledge the Center for Theoretical Underground Physics and Related Areas (CETUP* 2026) and the Institute for Underground Science at Sanford Underground Research Facility (SURF) for hospitality and for providing a conducive environment where this work was finalized.
CSF acknowledges the support by Fundacão de Amparo à Pesquisa do Estado de São Paulo (FAPESP) Contract No. 2019/11197-6  and Conselho Nacional de Desenvolvimento Científico e Tecnológico (CNPq) under Contract No. 304917/2023-0. SMCP acknowledges the support by FAPESP Contract No. 2024/17752-0.

\appendix

\section{Cosmic evolution of lepton flavor charges}
\label{app:lepton_flavor_charges}

For completeness, we collect here the relevant equations for the effective lepton-flavor-covariant formalism developed in ref.~\cite{Fong:2021xmi}.

The Boltzmann equation for $Y_{\tilde \Delta} \equiv Y_B I_{3\times 3}/3 - Y_{\Delta \ell}$ is
\begin{eqnarray}
	s{\cal H}z\frac{dY_{\tilde \Delta}}{dz} & = & \frac{\gamma_{E}}{2Y^{{\rm nor}}}\left\{ y_{E}^{\dagger}y_{E},\frac{Y_{\Delta\ell}}{g_{\ell}\zeta_{\ell}}\right\} - \frac{\gamma_{E}}{Y^{{\rm nor}}}y_{E}^{\dagger}y_{E}\frac{Y_{\Delta H}}{g_{H}\zeta_{H}} - \frac{\gamma_{E}}{Y^{{\rm nor}}}y_{E}^{\dagger}\frac{Y_{\Delta E}}{g_{E}\zeta_{E}}y_{E},
	\label{eq:BEtildeDelta}
\end{eqnarray}

while the Boltzmann equation for $Y_{\Delta E}$ is still given by eq.~\eqref{eq:BEE}.
When introducing the source and washout terms (for e.g. eqs.~\eqref{eq:source_typeI} and \eqref{eq:washout_typeI}, or eqs.~\eqref{eq:source_typeII} and \eqref{eq:washout_typeII}) to the equation above, they would come with an overall negative sign.
The quark Yukawa and EW sphaleron interactions are captured the following relations
\begin{eqnarray}
	Y_{\Delta \ell} &=& \frac{2}{15} c_B(T) I_{3\times 3} \,{\rm Tr} \,Y_{\tilde \Delta} - Y_{\tilde \Delta},\\
	Y_{\Delta H} &=& - c_H^{\rm eff}(T) \left({\rm Tr}\, Y_{\tilde \Delta}
	- 2\, {\rm Tr}\, Y_{\Delta E} 
	- 2 \sum_i q_{\phi_i}^Y Y_{\Delta \phi_i}  \right),
\end{eqnarray}
where $c_B(T) = 1 - e^{-T_B/T}$ with $T_B = 2.3 \times 10^{12}$ GeV and $c_H^{\rm eff}(T)$ is given by eq.~\eqref{eq:cH_eff}. 
The $B-L$ charge is given by
\begin{equation}
	Y_{B-L} = {\rm Tr} \,Y_{\tilde \Delta} - {\rm Tr} \,Y_{\Delta E}.
\end{equation}

\section{Description of the code}
\label{app:code}
For the numerical  implementation used in this work, we have constructed a Python package called \texttt{BOLEH}, which is available in \cite{CubidesPerez_BOLEH}. We have organized the formalisms into three independent modules corresponding to 
(i) the complete SM-flavor-covariant formalism, (ii) the effective quark-flavor-covariant formalism and (iii) the effective lepton-flavor-covariant formalism.

The main components of  each module are:
\begin{itemize}
    \item \textbf{physics$^*$.py}: contains the cosmological quantities, Standard Model parameters, and thermal functions, including the entropy density, Hubble expansion rate, and equilibrium quantities.
    
    \item \textbf{solver$^*$.py}: implements the numerical integration of the Boltzmann equations using the Backward Differentiation Formula (BDF) method.
    
    \item \textbf{interface$^*$.py}: handles the input parameters, post-processing of the numerical output, plotting routines, and communication between the different components of the package.
\end{itemize}

Each framework follows the same interface, which allows its application to new baryogenesis models. New physics models can be implemented by specifying the additional dynamical degrees of freedom with their corresponding Boltzmann equations, source and washout terms, and the model parameters. The SM evolution is then automatically combined with the equations defined by the user, providing as output the SM degrees of freedom evolution plots, together with the $B-L$ charge evolution and the numerical matrix results in a HDF5 file.   

The evolution variable of the package is 
\begin{equation}
    z=\frac{M_{\rm ref}}{T},
\end{equation}
where $M_{\rm ref}$ is a user-defined reference mass and the quark and lepton asymmetries are represented by $3\times3$ Hermitian matrices. The code maps these matrices into a vector state of independent varibles before integration and then reconstructs the matrix asymmetries  after each integration step.

By default, the code adopts the basis for quark Yukawa couplings $y_U = {\rm diag}(y_u,y_c,y_t)$, $y_D = V_{\rm CKM}{\rm diag}\,(y_d,y_s,y_b)$, using fixed values at various scales obtained from ref.~\cite{Antusch:2025fpm}
\begin{eqnarray}
	y_{t}\left(10^{16}\,{\rm GeV}\right) & = & 0.4454,\\
	y_{b}\left(10^{12}\,{\rm GeV}\right) & = & 0.719\times10^{-2},\\
	y_{c}\left(10^{9}\,{\rm GeV}\right) & = & 1.98\times10^{-3},\\
	y_{s}\left(10^{9}\,{\rm GeV}\right) & = & 1.72\times10^{-4},\\
	y_{d}\left(10^{7}\,{\rm GeV}\right) & = & 0.97\times10^{-5},\\
	y_{u}\left(10^{7}\,{\rm GeV}\right) & = & 4.39\times10^{-6},
\end{eqnarray}
and since the RGE running of the mixing angles and phase for the CKM mixing matrix $V_{\rm CKM}$ is negligible, we will use the values at $\mu=10^{12}\,{\rm GeV}$
\begin{eqnarray}
	\theta_{12} & = & 0.227,\\
	\theta_{23} & = & 4.65\times10^{-2},\\
	\theta_{13} & = & 4.11\times10^{-3},\\
	\delta & = & 1.139.
\end{eqnarray}
Moreover, the charged lepton Yukawa is taken to be the value at $\mu = 10^{12}$ GeV~\cite{Antusch:2025fpm}: 
\begin{equation}
	y_{E}={\rm diag}\left(2.8\times10^{-6},5.9\times10^{-4},1.0\times10^{-2}\right).
\end{equation}
These default values can be modified by the user whenever required.

Finally, the repository contains example implementations of the baryogenesis scenarios discussed in this work. We also prepare \textbf{example.py} to guide the users in implementing their own model.

\bibliography{biblio}

\end{document}